\documentclass[12pt]{article}

\usepackage{graphics}
\usepackage{epsfig}
\usepackage{cite}
\input epsf

\title{Inclusive $J/\psi$ photoproduction and polarization at HERA \\ in the $k_T$-factorization approach}

\author{S.P.~Baranov$^a$, A.V.~Lipatov$^b$, N.P.~Zotov$^b$}

\begin{document}

\maketitle

\begin{center}

{\it $^a$\,P.N.~Lebedev Physics Institute,\\ 
119991 Moscow, Russia\/}\\[3mm]

{\it $^b$\,D.V.~Skobeltsyn Institute of Nuclear Physics,\\ 
M.V. Lomonosov Moscow State University,
\\119991 Moscow, Russia\/}\\[3mm]

%{\it D.V.~Skobeltsyn Institute of Nuclear Physics,\\ 
%M.V. Lomonosov Moscow State University,
%\\119991 Moscow, Russia\/}\\[3mm]

\end{center}

\vspace{0.5cm}

\begin{center}

{\bf Abstract }

\end{center}

We investigate the inclusive photoproduction of $J/\psi$ mesons at HERA 
within the framework of the $k_T$-factorization QCD 
approach. Our consideration is based on the color singlet model supplemented
with the relevant off-shell matrix elements
and the CCFM and KMR unintegrated gluon densities in a proton and in a photon.
Both the direct and resolved photon contributions  
are taken into account.
Our predictions are compared with the recent
experimental data taken by the H1 and ZEUS collaborations. 
Special attention is put on the  $J/\psi$ polarization parameters
$\lambda$ and $\nu$ which are sensitive to the production dynamics.

\vspace{0.8cm}

\noindent
PACS number(s): 12.38.-t, 13.85.-t

\vspace{0.5cm}

\section{Introduction} \indent 

Heavy quarkonium production at high energies is subject of intense 
study from both theoretical and experimental points of view~\cite{1,2,3}.
The puzzling history traces back to the early 1990s, when the measurements of 
the $J/\psi$ and $\Upsilon$ hadroproduction cross sections
at  Tevatron energies revealed a more 
than one order-of-magnitude discrepancy with the theoretical expectations 
of the color singlet (CS) model~\cite{4}.
This fact has induced extensive theoretical activity, mainly in the
description of the formation of bound $q \bar q$ state from the heavy quark pair produced
in the hard interation (photon-gluon or gluon-gluon fusion). 
In the CS model, only those states 
with the same quantum numbers as the resulting quarkonium contribute to the 
formation of a bound state. This is achieved by radiating a hard gluon in a
perturbative process. 
In the color octet (CO) model~\cite{5}, it was suggested to 
add the contribution of transition mechanism from $q\bar q$ pairs to quarkonium, 
where a heavy quark pair is produced in a color octet 
state and transforms into the final color singlet state by the help of 
soft gluon radiation. 
The CO model is based on the general principle of 
the non-relativistic QCD factorization (NRQCD)~\cite{6}.
By adding the contribution from the octet states and fitting the free parameters 
one was able to describe the existing data on the $J/\psi$ and $\Upsilon$ 
production at the Tevatron~\cite{7}.

However, recently the NLO QCD corrections to CS mechanism have been found 
to be essential in quarkonium production~\cite{8}.
The large discrepances between experiments and LO CS predictions 
for $J/\psi$ and $\Upsilon$ inclusive production can be largely resolved at NLO level 
within the CS model~\cite{8, 9}. These NLO corrections 
to the $J/\psi$ production at the Tevatron enhance the cross section at 
large $p_T$ by two orders of magnitude~\cite{9, 10}. 
The complete NLO calculations of the $J/\psi$ photoproduction at HERA
slighly favor the presence of CO contributions~\cite{11, 12}. 
In order to achive agreement between the last calculations and the
experimental data for the $p_T$-distributions measured at HERA and Tevatron 
the contribution of CO P-wave
state to be asummed is negative~\cite{12, 13}. 
%  There are also some another 
%indications that CO model does not work well. For example, to describe the 
%$J/\psi$ photoproduction data at HERA (namely, distributions in elasticity $z$) 
%the contribution from the CO mechanism is unnecessary or even unwanted since the 
%experimental results can be reasonable well reproduced~[11] within the 
%CS model alone (if the NLO contributions are taken into account).
Another problem refers to the quarkonium spin alignment: the CO model predicts 
the strong transverse polarization of the produced mesons (see~\cite{1}). 
This is in disagreement with the data which point to unpolarized or even 
longitudinally polarized quarkoniums~\cite{14, 15}.

A different strategy is represented by the $k_T$-factorization approach of QCD~\cite{16}.
This approach is based on the Balitsky-Fadin-Kuraev-Lipatov (BFKL)~\cite{17} or 
Ciafaloni-Catani-Fiorani-Marchesini (CCFM)~\cite{18} evolution equations 
for non-collinear gluon densities in a proton or photon.  A detailed description and discussion of 
the $k_T$-factorization approach can be found, for example, in reviews~\cite{19}. 
Quarkonium production and polarization have own story in the $k_T$-factorization approach.
Shortly, it was demonstrated~\cite{20, 21, 22, 23, 24} that in the framework of this approach the 
experimental data on the heavy quarkonim
production at HERA, RHIC and Tevatron can be reasonable well described 
without the CO contributions. It is important also that the kt-factorization 
predicts the longitudinal polarization of $J/\psi$ and $\Upsilon$ mesons
as an immediate consequence of initial gluon off-shellness~\cite{25}.
The results of recent theoretical caclulations
of the NLO and NNLO corrections to CS quarkonium production in the framework of 
standard pQCD~\cite{8, 9} are in much better
agreement with the $k_T$-factorization predictions~\cite{20, 21, 22, 23, 24} than it
was seen for LO collinear calculations.

The present note is motivated by the very recent experimental measurements~\cite{26, 27}
of the inclusive $J/\psi$ photoproduction cross sections performed by the H1 and ZEUS 
collaborations at HERA. These measurements have been
performed with increased statistics and precision as compared with 
previous experimental analyses~\cite{28, 29}. The number of single and double differential 
cross sections are determined and the polarisation parameters 
$\lambda$ and $\nu$ are measured in the several different reference frames
(namely helicity, target and Collins-Soper frames) as a function of 
$J/\psi$ transverse momentum and elasticity variable $z$.
Our main goal is to give a systematic analysis of available 
experimental data~\cite{26, 27, 28, 29} in the
framework of the CS model and the $k_T$-factorization approach supplemented with the 
CCFM gluon dynamics, what we not done in our first paper~\cite{20}.
In this analysis we will take into account 
both direct and resolved photon
contributions (last of them are important at low $z$).
Specially we concentrate on the $J/\psi$ polarization observables since
it can be useful, in particular, in discriminating the CS 
and CO production mechanisms.
Also we will study the possible sources of theoretical uncertainties 
of our predictions (i.e. uncertainties connected with the gluon evolution scheme
and hard scale of partonic subprocess). 
To investigate the dependence of our predictions on 
the non-collinear evolution scheme we will apply the unintegrated 
gluon densities derived from the usual (i.e. DGLAP-evolved) parton 
distributions (in the Kimber-Martin-Ryskin (KMR)~\cite{30} 
approximation). 

%The similar calculations were
%made very recently in~[18] and a reasonable well description
%of the H1 data~[4] has been obtained. However, the ZEUS data~[1, 2]
%have not been analysed still using the CCFM-evolved gluon densities. Our
%calculations %have been performed independently from~[18] and
%covers all available (up to now) experimental data.

The outline of our paper is following. In Section~2 we 
recall shortly the basic formulas of the $k_T$-factorization approach with a brief 
review of calculation steps. In Section~3 we present the numerical results
of our calculations and a discussion. Section~4 contains our conclusions.

\section{Theoretical framework} \indent 

In the present note we follow the approach described in detail 
in our previous paper~\cite{20}. Here we only briefly recall the basic
formulas.

In the framework of the CS model, the production of any heavy quarkonium
is described as the perturbative production of a color singlet quark-antiquark pair 
in a state with properly arranged the quantum numbers, according to the quarkonium 
state under consideration. Our calculations in the $k_T$-factorization approach 
are based on 
the off-shell ($k_T$-dependent) photon-gluon and gluon-gluon fusion 
subprocesses $\gamma g^* \to J/\psi g$ and 
$g^* g^* \to J/\psi g$. The spin projection operator~\cite{4} 
$$
  J(p_\psi) = \hat \epsilon_{\psi} (\hat p_{\psi} + m_{\psi})/2 m_{\psi}, \eqno(1)
$$

\noindent 
is included to guarantee the proper quantum numbers of the created charmed quark pair.
Here $m_{\psi}$, $p_\psi$ and $\epsilon_{\psi}$ are the mass, four-momentum and
polarization vectors of produced $J/\psi$.
In accordance with the non-relativistic formalism of bound state formation, the charmed
quarks are assumed to each carry one half of the $J/\psi$ momentum and the
probability of creation of $J/\psi$ meson is determined by a value of wave function 
at the origin of coordinate space $|\Psi(0)|^2$, which is known from the
leptonic decay width~\cite{31}.
The calculation of relevant off-shell matrix elements 
$|\bar {\cal M}(\gamma g^* \to J/\psi g)|^2$ and $|\bar {\cal M}(g^* g^* \to J/\psi g)|^2$
has been described detally in~\cite{20}.
Here we would like to only mention two technical points. First,
in according to the $k_T$-factorization prescription~\cite{16},
the summation over the incoming off-shell gluon polarizations is 
carried with $\sum \epsilon^\mu \epsilon^{*\, \nu} = {\mathbf k}_T^{\mu} {\mathbf k}_T^{\nu}/{\mathbf k}_T^2$,
where ${\mathbf k}_T$ is the gluon transverse momentum.
Second, the spin density matrix of $J/\psi$ meson is determined by the
momenta $l_1$ and $l_2$ of the decay leptons and is taken in the 
form\footnote{This formula has misprint in~\cite{22}.}
$$
  \sum \epsilon^\mu_{\psi} \epsilon^{*\, \nu}_{\psi} = 3 \left( l_1^\mu l_2^\nu + l_1^\nu l_2^\mu - {m_{\psi}^2\over 2} g^{\mu \nu} \right)/m_\psi^2. \eqno(2)
$$

\noindent 
This expression is equivalent to the standard one 
$\sum \epsilon^\mu_{\psi} \epsilon^{*\, \nu}_{\psi} = - g^{\mu \nu} + p_\psi^\mu p_\psi^\nu/m_\psi^2$ 
(which has been used previously in~\cite{20, 21}) but is better suited for 
studying the polarization variables.

The cross section of $J/\psi$ photoproduction 
at high energies in the $k_T$-factorization approach
is calculated as a convolution of the off-shell 
partonic cross section $\hat \sigma$ and the unintegrated gluon 
distributions in a proton and in a photon. The direct and resolved photon 
contributions can be presented in the following form:
$$
  \displaystyle \sigma_{\rm dir} (\gamma p \to J/\psi \, X) = \int {1\over 16\pi (x_2 W^2)^2 } {1\over z(1 - z)} {\cal A}(x_2,{\mathbf k}_{2T}^2,\mu^2) \times \atop
  \displaystyle  \times |\bar {\cal M}(\gamma g^* \to J/\psi g)|^2 d{\mathbf p}_{\psi\,T}^2 d{\mathbf k}_{2T}^2 dz {d\phi_2 \over 2\pi}, \eqno (3)
$$
$$
  \displaystyle \sigma_{\rm res} (\gamma p \to J/\psi \, X) = \int {1\over 16\pi (x_1 x_2 W^2)^2 } {\cal A}_\gamma(x_1,{\mathbf k}_{1T}^2,\mu^2) {\cal A}(x_2,{\mathbf k}_{2T}^2,\mu^2) \times \atop
  \displaystyle  \times |\bar {\cal M}(g^* g^* \to J/\psi g)|^2 d{\mathbf p}_{\psi\,T}^2 d{\mathbf k}_{1T}^2 d{\mathbf k}_{2T}^2 dy_\psi dy_g {d\phi_1 \over 2\pi} {d\phi_2 \over 2\pi}, \eqno (4)
$$

\noindent
where ${\cal A}(x,{\mathbf k}_{T}^2,\mu^2)$ and 
${\cal A}_\gamma(x,{\mathbf k}_{T}^2,\mu^2)$ are the
unintegrated gluon distributions in a proton and in a photon,
$y_\psi$ and $y_g$ are the rapidities of produced
$J/\psi$ meson and outgoing gluon and $W$ is the photon-proton center-of-mass energy.
The initial off-shell gluons have a fraction $x_1$ and $x_2$ 
of the parent proton and photon longitudinal 
momenta, non-zero transverse momenta ${\mathbf k}_{1T}$ and 
${\mathbf k}_{2T}$ (${\mathbf k}_{1T}^2 = - k_{1T}^2 \neq 0$, 
${\mathbf k}_{2T}^2 = - k_{2T}^2 \neq 0$) and azimuthal angles
 $\phi_1$ and $\phi_2$. 
The elasticity $z$ denotes the fractional energy of the photon 
transferred to the $J/\psi$ meson in the proton rest system:
$z = (p_\psi \cdot p_p)/(p_\gamma \cdot p_p)$, where
$p_\gamma$ and $p_p$ are the four-momenta of the incoming photon and proton.

The unintegrated gluon distributions in a 
proton and in a photon involved in~(3) and (4) can be obtained from the analytical or 
numerical solutions of the BFKL or CCFM
evolution equations. In the numerical calculations we have tested 
a few different sets. First of them, unintegrated gluon density in a proton
(CCFM set A0) has been obtained in~\cite{32} from the CCFM equation
where all input parameters have been fitted to describe the proton structure function $F_2(x, Q^2)$.
Equally good fit was obtained using different values for the soft cut 
and a different value for the width of the intrinsic ${\mathbf k}_{T}$ distribution 
(CCFM set B0). A reasonable description of the $F_2$ data
can be achieved~\cite{32} by both A0 and B0 sets. 
These unintegrated gluon densities in a proton will be supplemented
with the CCFM-evolved gluon density in a photon proposed in~\cite{33}.
Also we will use the unintegrated gluon densities in a proton and in a photon
taken in the Kimber-Martin-Ryskin form~\cite{30}. The KMR approach is a formalism to construct the 
unintegrated parton distributions from well-known conventional ones. 
For the input, we have used the leading-order GRV parametrizations~\cite{34}.

The multidimensional integrations in~(3) and~(4) have been performed
by the means of Monte Carlo technique, using the routine \textsc{vegas}~\cite{35}.
The full C$++$ code is available from the authors on 
request\footnote{lipatov@theory.sinp.msu.ru}.

\section{Numerical results} \indent

We now are in a position to present our numerical results. First we describe our
input and the kinematic conditions. After we fixed the unintegrated
gluon distributions, the cross sections (3) and (4) depend on
the renormalization and factorization scales $\mu_R$ and $\mu_F$. 
In the numerical calculations we set 
$\mu_R = \xi \sqrt{m_\psi^2 + {\mathbf p}_{\psi\,T}^2}$,
$\mu_F = \xi \sqrt{\hat s + {\mathbf Q}_T^2}$, where ${\mathbf Q}_T$ is the 
transverse momentum of initial off-shell gluon or gluon pair (in the case of
resolved photon production). In order to estimate the theoretical 
uncertainties of our calculations we vary the scale parameter 
$\xi$ between 1/2 and 2 about the default value $\xi = 1$.
The sensitivity of the predictions to the charmed quark mass
has been investigated previously in~\cite{20, 21}. Here we set
$m_c = 1.5$~GeV and use the LO formula 
for the coupling constant $\alpha_s(\mu^2)$ with $n_f = 4$ quark flavours
at $\Lambda_{\rm QCD} = 200$~MeV, such that $\alpha_s(M_Z^2) = 0.1232$. 
Note that we apply another value $\Lambda_{\rm QCD} = 220$~MeV for the CCFM-evolved
gluon densities (see discussion in Sect.~3.1).
Finally, the $J/\psi$ wave function at the origin
of coordinate space is taken to be equal to $|\Psi(0)|^2 = 0.0876$~GeV$^3$~\cite{31}.

\subsection{Inclusive production} \indent

Experimental data for the inclusive $J/\psi$ photoproduction at HERA
come from both the ZEUS and H1 collaborations. 
The dependence of the total cross section on the photon-proton center-of-mass
energy $W$ has been measured and the number of single and double differential 
cross sections has been determined: $d\sigma/dz$, 
$d\sigma/d{\mathbf p}_{\psi\,T}^2$, $d\sigma/dy_{\psi}$ and 
$d\sigma/d{\mathbf p}_{\psi\,T}^2 dz$.
The ZEUS data~\cite{29} refer to the kinematic region
defined by $0.1 < z < 0.9$, ${\mathbf p}_{\psi\,T}^2 > 1$~GeV$^2$ and 
$50 < W < 180$~GeV.
The H1 data~\cite{28} refer to the kinematic region
$0.05 < z < 0.9$, $1 < {\mathbf p}_{\psi\,T}^2 < 60$~GeV$^2$ and 
$60 < W < 260$~GeV. Very recent H1 measurements~\cite{27} have been performed
at $0.3 < z < 0.9$, $1 < {\mathbf p}_{\psi\,T}^2 < 100$~GeV$^2$ and
$60 < W < 240$~GeV.

Our numerical predictions are shown in Figs.~1 --- 7 in 
comparison with the data~\cite{27, 28, 29}. The solid, dotted and dash-dotted
histograms correspond to the results obtained using the 
CCFM A0, B0 and KMR gluon densities, respectively.
The upper and lower dashed histograms represent the scale 
scale variations as it was described above.
To estimate this uncertainty we used 
the CCFM set A0$+$ and A0$-$ instead of the default gluon density A0: 
the A0$-$ stands for $\xi = 1/2$ while set A0$+$ reflects $\xi = 2$. 
One can see a good overall agreement of our central CCFM predictions 
with the recent H1 data~\cite{27}. We would like to point out that the shape of $z$ 
distributions is well reproduced by the CS mechanism.
It is in a contrast with the predictions~\cite{12} where the octet 
contributions have been taken into account.
The transverse momentum distributions and 
the dependence of $J/\psi$ total cross section on the
energy $W$ are well described also.
It seems that the agreement of our CCFM predictions with the previous H1 data~\cite{28} 
is slightly less satisfactory but still reasonable
(see Figs.~4 --- 6). Main difference between data and theory occurs in the high 
${\mathbf p}_{\psi\,T}$ region.
We note, however, that the measurements~\cite{27, 28} have been
performed in essentially the same kinematical region and therefore probably 
there exists a problem of incompatibility of these data with each other.
The shape of $y_{\psi}$ distribution measured by ZEUS is not 
fully reproduced, but the data points lie still within the 
band of theoretical uncertainties. In general, the overall agreement of our 
CCFM predictions with ZEUS data~\cite{29} is reasonable well (see Fig.~7).
The KMR gluon density is also responsible for description of the H1 and 
ZEUS data although the shape of $W$-dependence of $J/\psi$ total 
cross section is not reproduced well.

The predictions of the Monte Carlo event generator \textsc{Cascade}~\cite{36}
have been obtained in the kinematical range of recent H1 measurements~\cite{27}.
\textsc{Cascade} is a full hadron level event generator which uses the CCFM 
equation for the initial state gluon evolution.
Despite the fact that the default sets of numerical parameters
are the same, the \textsc{Cascade} predictions (based on the A0 gluon
density) lie somewhat above than ours. 
The possible reason can be connected with the corrections for contributions from 
the non-resonant background events which have been included into the
\textsc{Cascade} calculations (see more details in~\cite{27}).
These corrections are not taken into account in our consideration.
In addition, there are some effects from the final-state parton showers 
in \textsc{Cascade} which are also absent in our approach.
However, we obtain the same results as coming from \textsc{Cascade}
if we apply the small shift in QCD parameter to $\Lambda_{\rm QCD} = 220$~MeV.
This is the reason why such shifted value of $\Lambda_{\rm QCD}$ 
has been used in our CCFM calculations. 
In this way we effectively simulate the missing final-state parton showers effects and 
effects coming from non-resonant background events.
It leads to increase of the predicted cross sections by the 15\% approximately,
that is, of course, much less than the 
scale uncertainties of our calculations.
Note that the choice $\Lambda_{\rm QCD} = 200$~MeV in the KMR predictions
is defined by the set of relevant parameters in GRV parametrizations.

\subsection{Polarization properties} \indent

Now we turn now to the investigation of
the $J/\psi$ polarization in photoproduction events at HERA.
As it was mentioned above, the polarization observables are useful
in discriminating the CS and CO production mechanisms, as well 
as collinear and $k_T$-factorization approaches.
In general, the spin density matrix of a vector
particle depends on three parameters $\lambda$, $\mu$ and $\nu$ which can be
measured experimentally from the double differential
angular distribution of the decay products.
The latter reads for the $J/\psi$ leptonic decay~\cite{37}
$$
  {d\sigma \over d\phi^* d\cos \theta^*} \sim 1 + \lambda \cos^2 \theta^* + 
    \mu \sin 2 \theta^* \cos \phi^* + {\nu \over 2} \sin^2 \theta^* \cos 2 \phi^*, \eqno(5)
$$

\noindent
where $\theta^*$ and $\phi^*$ are the polar and azimuthal angles
of the decay lepton measured in the $J/\psi$ rest frame.
In the H1 experiments~\cite{27, 28} the polarization
parameters $\lambda$ and $\nu$ have been measured as a functions
of ${\mathbf p}_{\psi\,T}$ and $z$ in two complementary
frames: the helicity frame and Collins-Soper frame.
In addition, the ZEUS collaboration performs the measurements in the target frame~\cite{26, 29}.
In the helicity frame the polarization axis in the $J/\psi$ meson 
rest frame is defined by the flight direction of the $J/\psi$ 
meson in the $\gamma p$ rest frame, whereas the polarization in the 
Collins-Soper frame is measured with respect to the bisector of 
proton ($ - \vec p_p$) and photon ($\vec p_\gamma$) 
in the $J/\psi$ meson rest frame.
In the target frame the polarization axis is chosen to be opposite of the 
incoming proton direction in the $J/\psi$ rest frame.
Note that cases $\lambda = 1$ and $\lambda = - 1$ correspond to transverse and 
longitudinal polarization of the $J/\psi$ meson, respectively.
Unfortunately, the experimental statistics
was insufficient to encourage the extraction of
parameter $\mu$ from the double differential cross section~(5).
Our theoretical calculations were adjusted to the
experimental binning and generally followed the experimental
procedure. We have collected the simulated
events in the specified bins of $p_{\psi\,T}$ and $z$,
generated the decay lepton angular distributions according
to the production and decay matrix elements,
and then applied a three-parametric fit based on~(5).

The estimated values of $\lambda$ and $\nu$ are shown
in Figs.~8 --- 15 in comparison with the H1~\cite{27, 28} and ZEUS experimental data~\cite{26, 29}.
The solid and dash-dotted histograms represent the results obtained 
in the $k_T$-factorization approach with the CCFM A0 and KMR gluon densities.
The dotted histograms correspond to the LO CS model predictions.
First of all, one can clearly see the difference between the collinear 
and $k_T$-factorization approaches in behavior of 
parameter $\lambda$ as a function of $p_{\psi\,T}$.
In contrast with the LO CS model, the $k_T$-factorization 
predicts the longitudinal polarization of produced $J/\psi$ mesons 
at high transverse momenta. Note that if the NLO corrections to 
the LO CS cross section will be 
taken into account, the $J/\psi$ polarization is also tends to be
longitudinal~\cite{8}. It is not the case of CO predictions 
where NLO corrections keep the essential transverse polarization.
The parameter $\lambda$ as a function of $z$ seems to be insufficient
to discriminate between the two production mechanisms. Contrary, behaviour of 
parameter $\nu$ as a function of $z$ demonstrates the analyzing power
and can be useful to distiguish both the approaches.

Concerning the comparison of our predictions and the data,
one can see that none of the predictions can describe all aspects of the
data even though huge experimental uncertainties.
However, the kt-factorization approach tends to be in a better agreement
with the data, especially with respect to the $p_{\psi\, T}$ dependence 
of $\lambda$ and $z$ dependence of $\nu$. 
We point out that our predictions for the polarization parameters 
$\lambda$ and $\nu$ are not sensitive to the unintegrated gluon density used.
So one of the sources of theoretical uncertainties is cancels out.
Therefore measurements of the polarization parameters can to play
crucial role in discriminating the different theoretical
approaches.

\section{Conclusions} \indent 

We have studied inclusive $J/\psi$ meson photoproduction 
at HERA within the framework of the $k_T$-factorization approach.
Our consideration is based on the usual CS model supplemented
with the relevant off-shell matrix elements
and the CCFM-evolved unintegrated gluon densities in a proton and in a photon.
Both the direct and resolved photon contributions  
are taken into account.
The analysis covers the total, single and double differential cross sections 
of $J/\psi$ mesons. Special attention was put on the
polarization parameters $\lambda$ and $\nu$ which determine the 
$J/\psi$ spin density matrix.

We obtained a reasonable well agreement of our calculations
and the recent experimental data taken by the H1 and ZEUS collaborations.
We demonstrated that the $k_T$-factorization approach 
gives a better agreement with the polarization data than the 
leading-order collinear calculations.
Both polarization parameters $\lambda$ and $\nu$ can be used 
to discriminate the $J/\psi$ production mechanism
and study the parton interaction dynamics.

\section{Acknowledgements} \indent 

%We thank S.P.~Baranov for his encouraging interest, useful discussions
%and help in the numerical calculations.
The authors are very grateful to 
DESY Directorate for the support in the 
framework of Moscow --- DESY project on Monte-Carlo
implementation for HERA --- LHC.
A.V.L. was supported in part by the Helmholtz --- Russia
Joint Research Group.
Also this research was supported by the 
FASI of Russian Federation (grant NS-4142.2010.2) and 
FASI state contract 02.740.11.0244.

\newpage

\begin{figure}
\begin{center}
\epsfig{figure=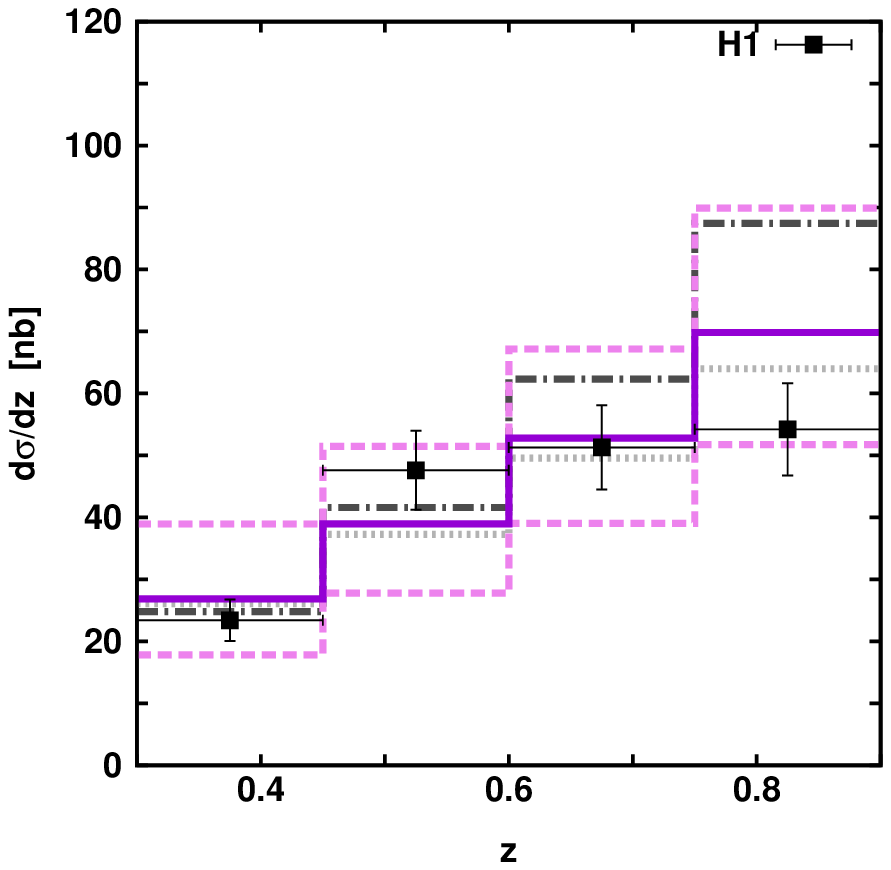, width = 8.1cm}
\epsfig{figure=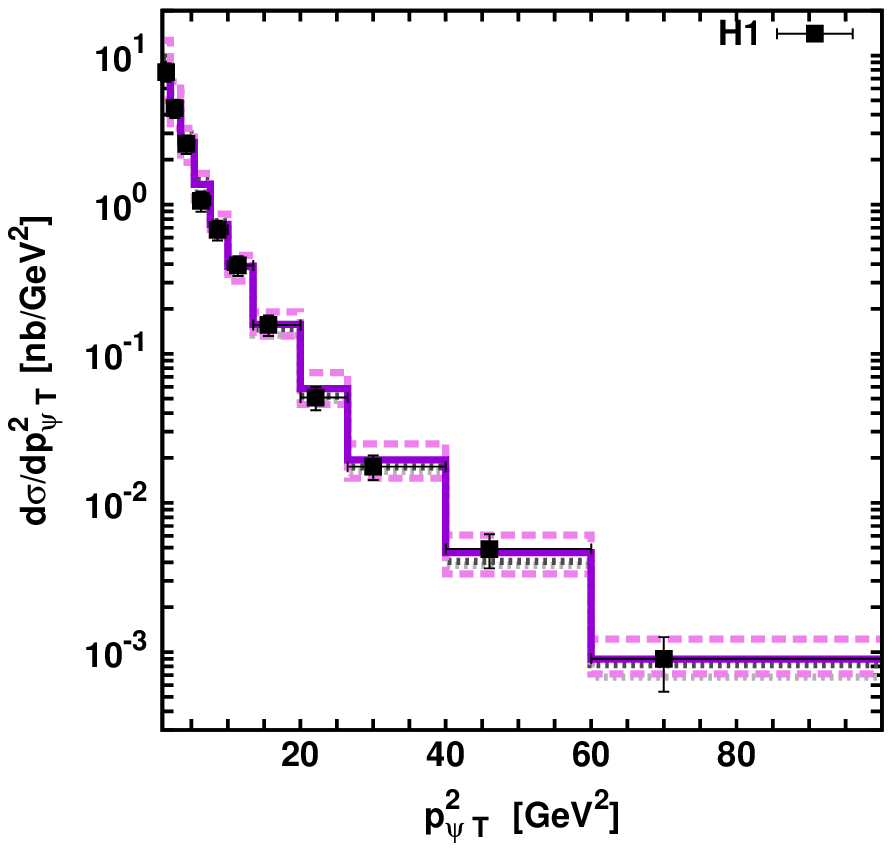, width = 8.1cm}
\epsfig{figure=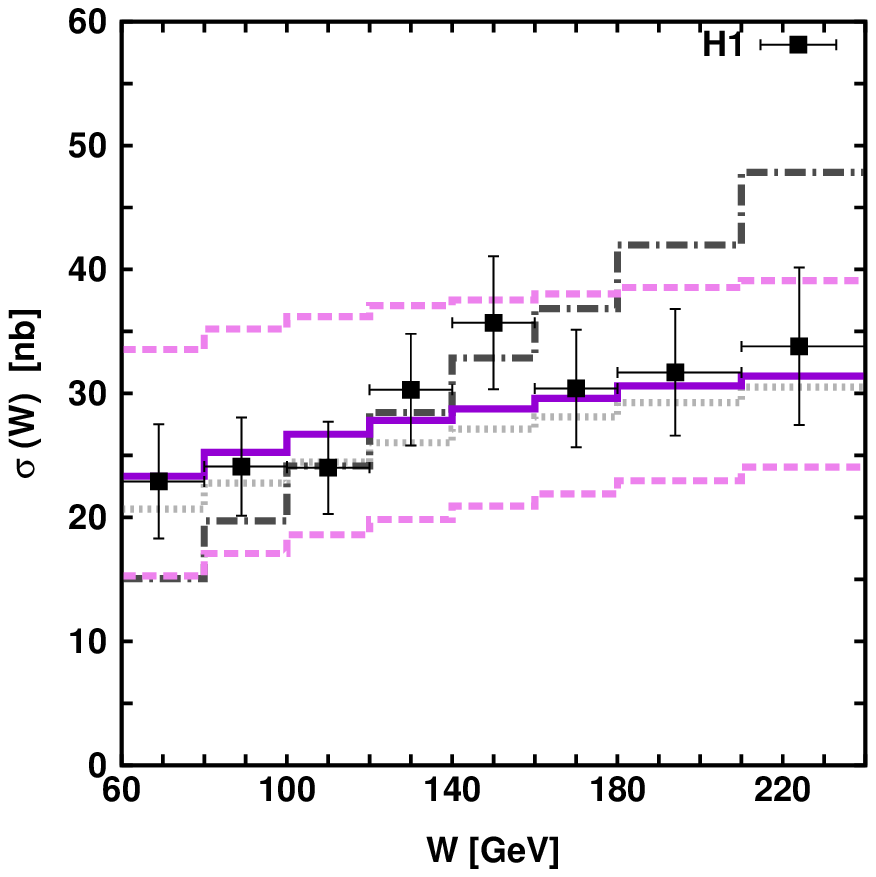, width = 8.1cm}
\caption{The total and differential cross sections of $J/\psi$ mesons calculated
in the kinematical region defined by $0.3 < z < 0.9$, 
${\mathbf p}_{\psi\,T}^2 > 1$~GeV$^2$ and $60 < W < 240$~GeV.
The solid, dotted and dash-dotted
histograms correspond to the results obtained using the 
CCFM A0, CCFM B0 and KMR gluon densities, respectively.
The upper and lower dashed histograms represent the scale 
scale variations as it is described in the text.
The experimental data are from H1~\cite{27}.}
\end{center}
\label{fig1}
\end{figure}

\newpage

\begin{figure}
\begin{center}
\epsfig{figure=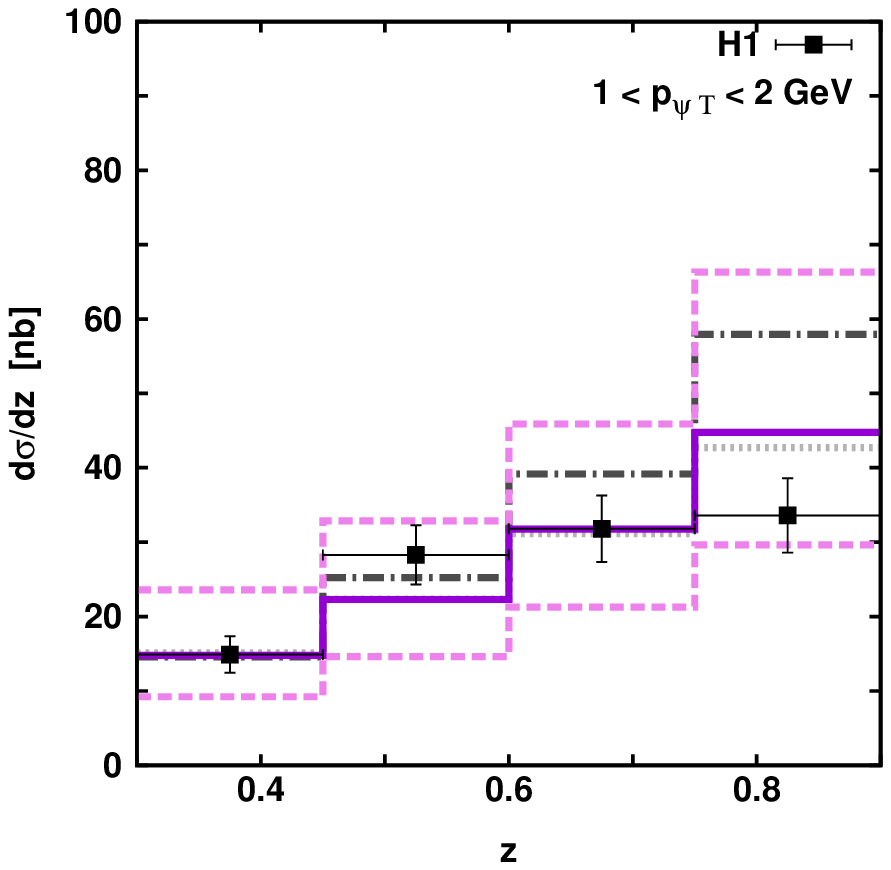, width = 8.1cm}
\epsfig{figure=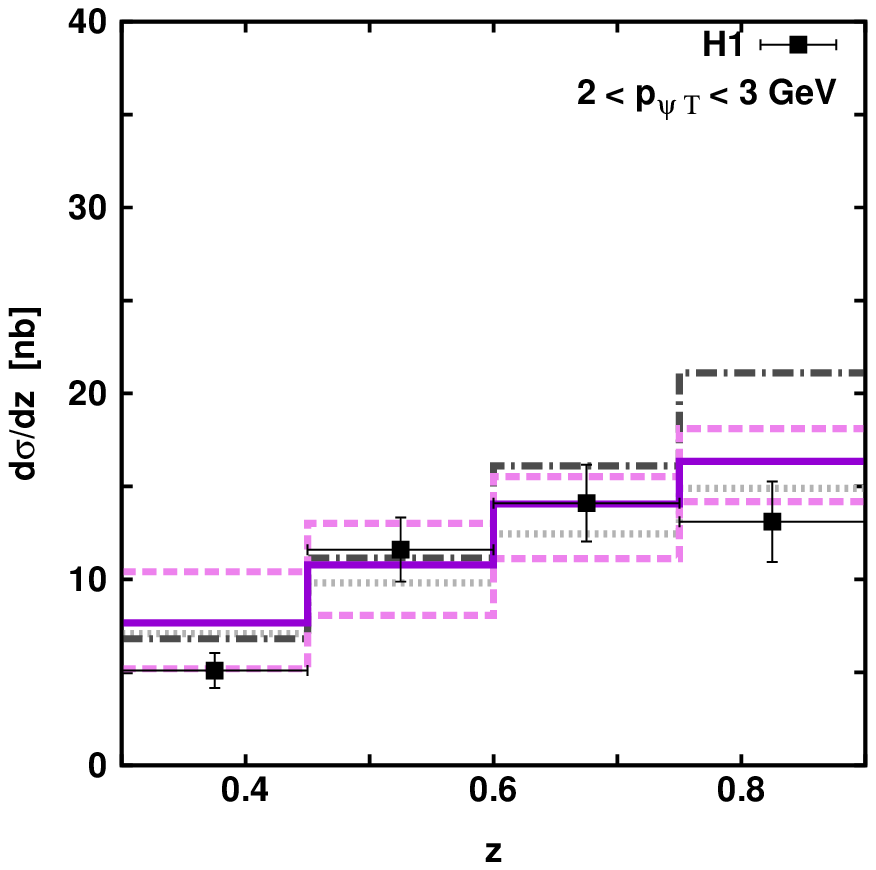, width = 8.1cm}
\epsfig{figure=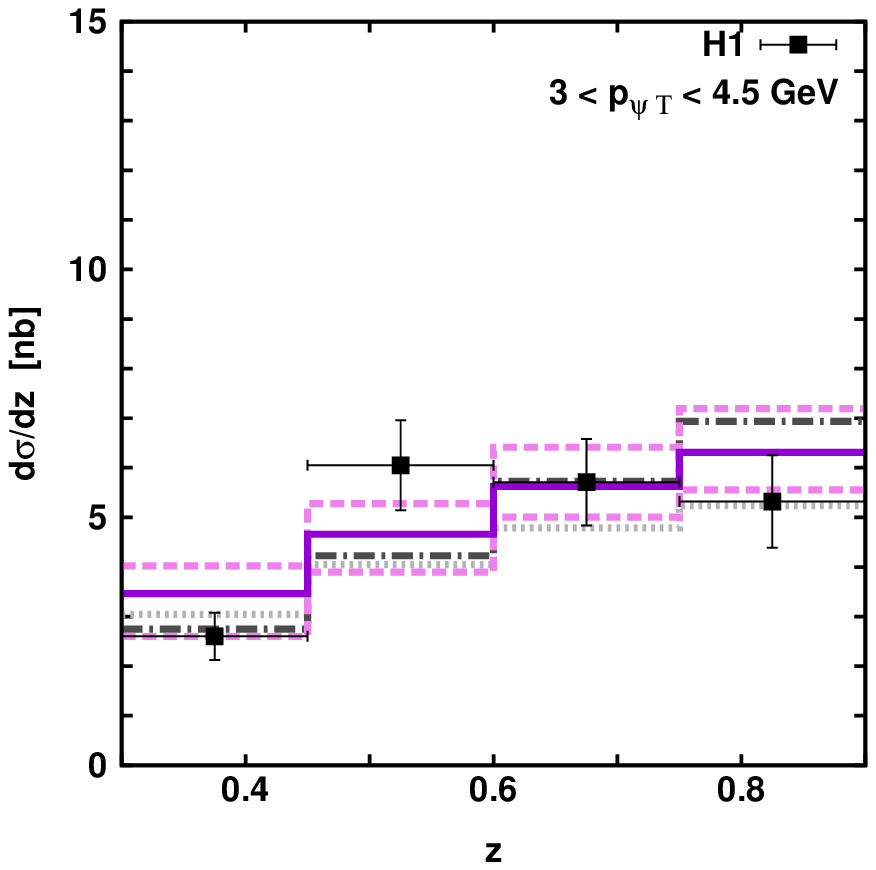, width = 8.1cm}
\epsfig{figure=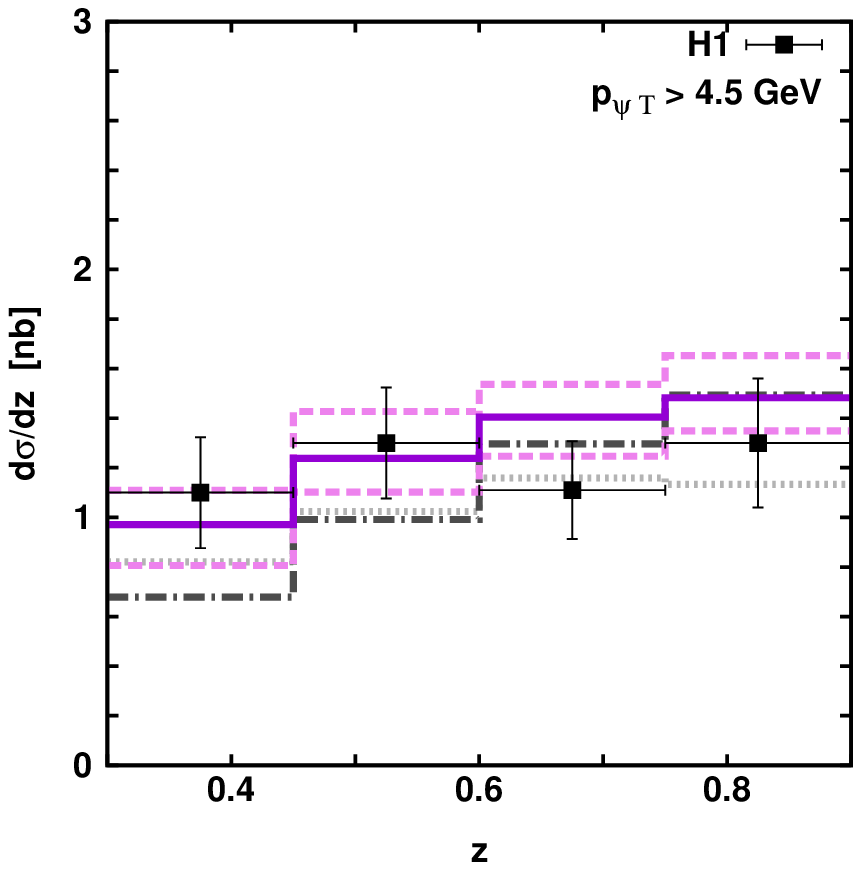, width = 8.1cm}
\caption{The differential cross sections of $J/\psi$ mesons as a function of elasticity $z$ 
calculated in different ${\mathbf p}_{\psi \,T}$ bins at $60 < W < 240$~GeV. 
Notation of all histograms is the same as in Fig.~1.
The experimental data are from H1~\cite{27}.}
\end{center}
\label{fig2}
\end{figure}

\newpage

\begin{figure}
\begin{center}
\epsfig{figure=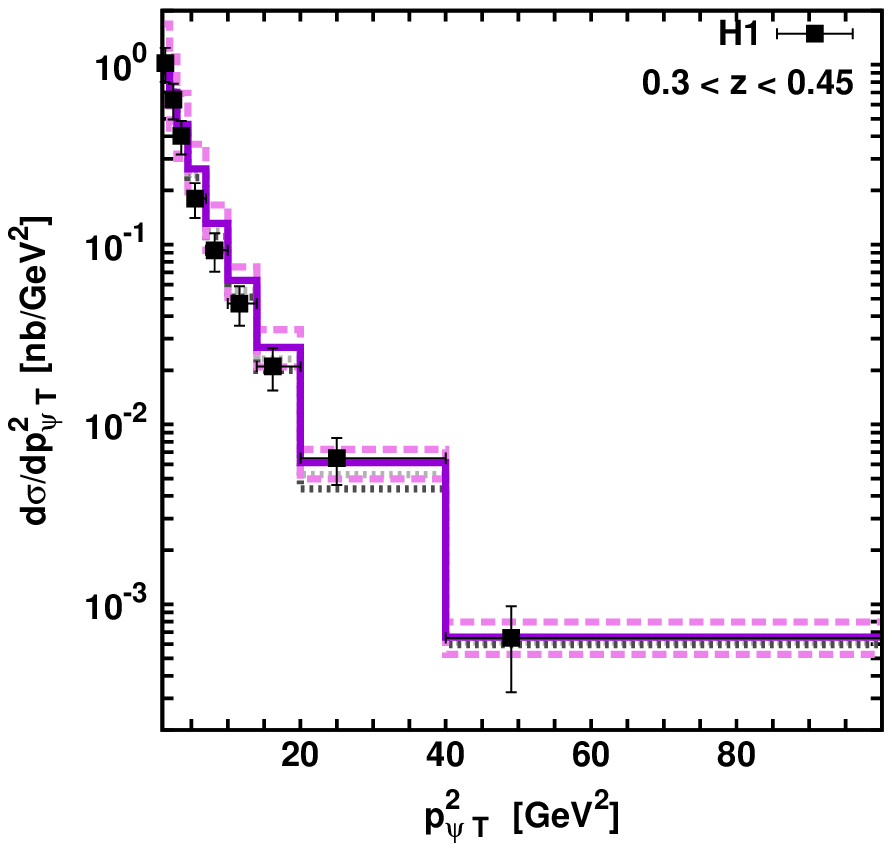, width = 8.1cm}
\epsfig{figure=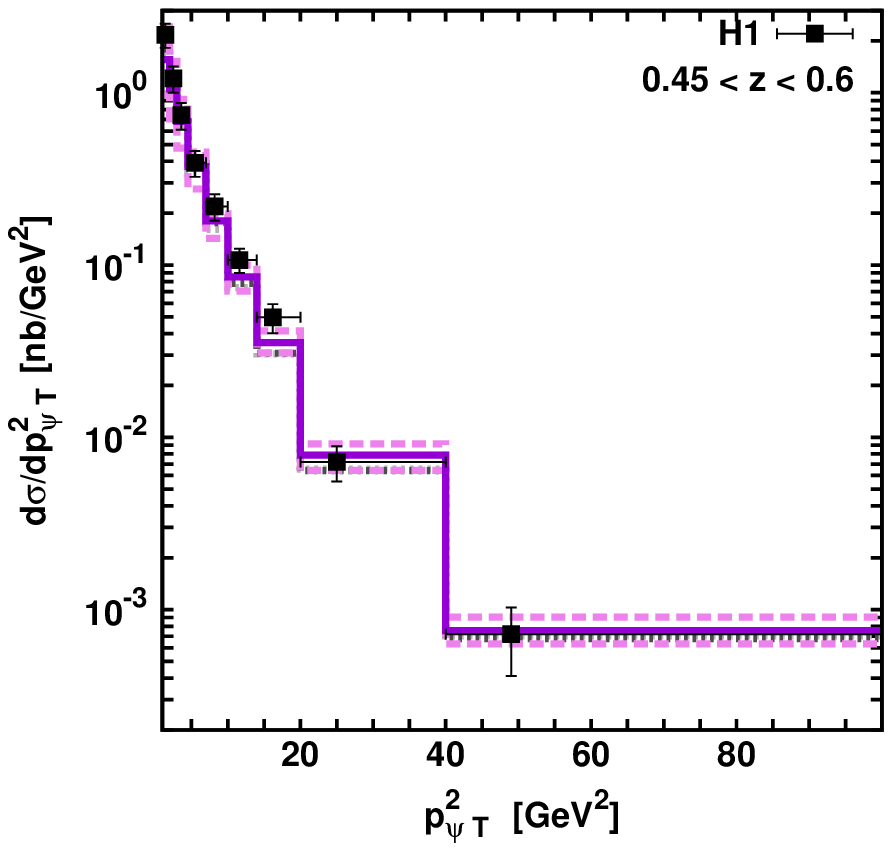, width = 8.1cm}
\epsfig{figure=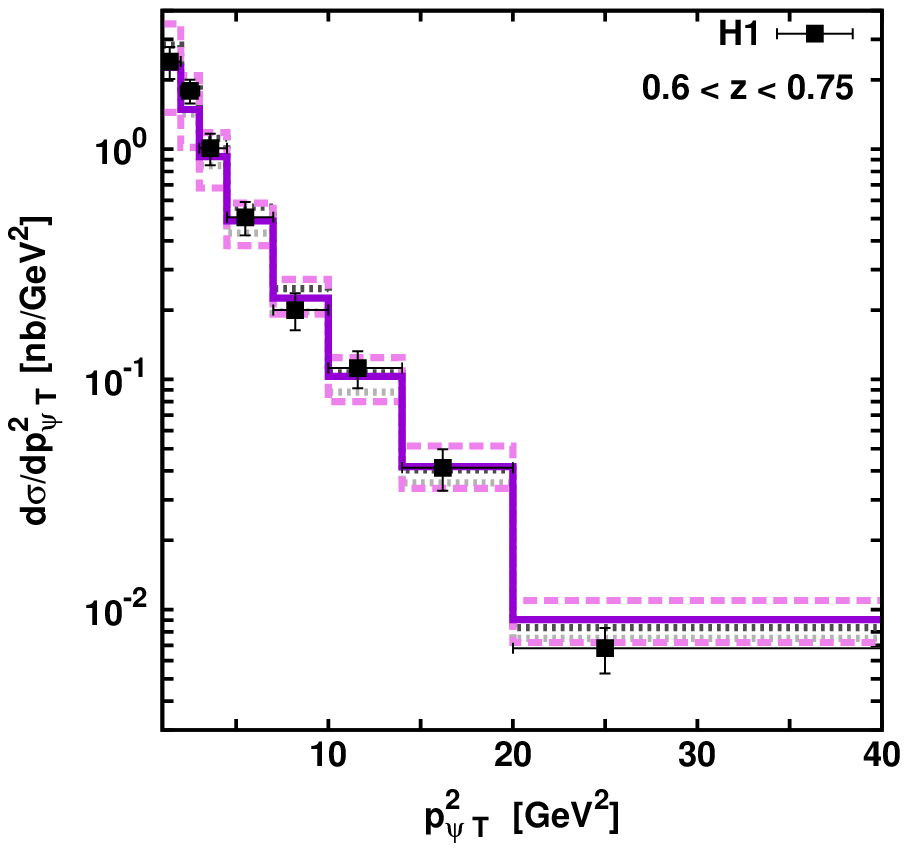, width = 8.1cm}
\epsfig{figure=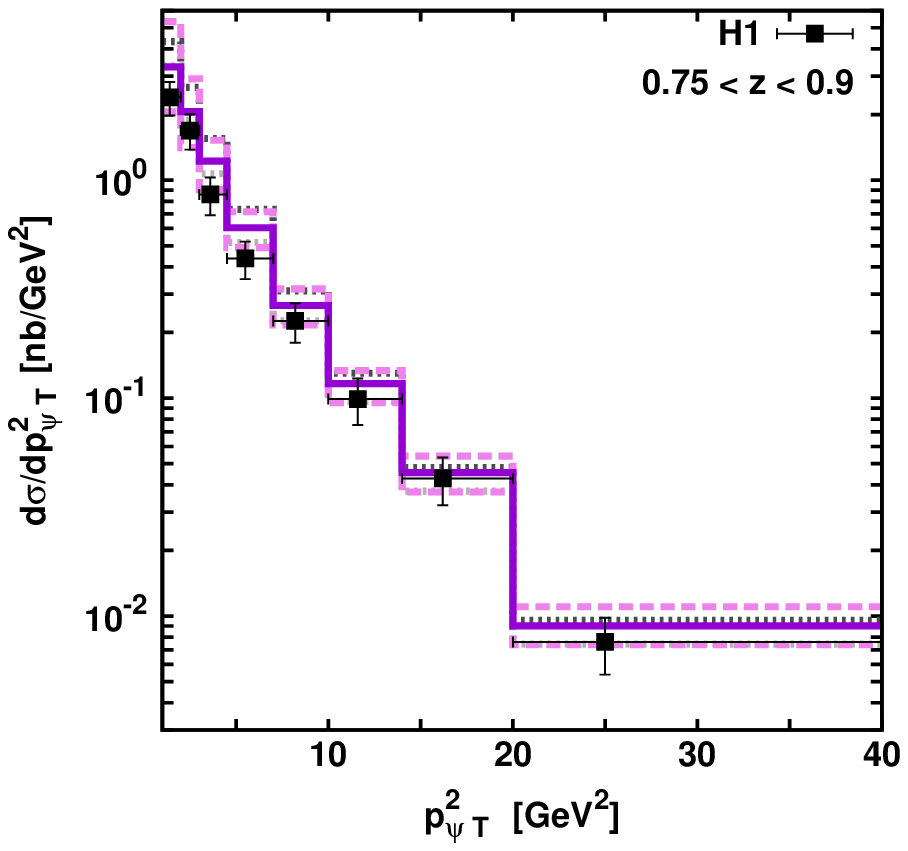, width = 8.1cm}
\caption{The differential cross sections of $J/\psi$ mesons as a function 
of ${\mathbf p}_{\psi \,T}^2$ calculated in 
different $z$ bins at $60 < W < 240$~GeV. 
Notation of all histograms is the same as in Fig.~1.
The experimental data are from H1~\cite{27}.}
\end{center}
\label{fig3}
\end{figure}

\newpage

\begin{figure}
\begin{center}
\epsfig{figure=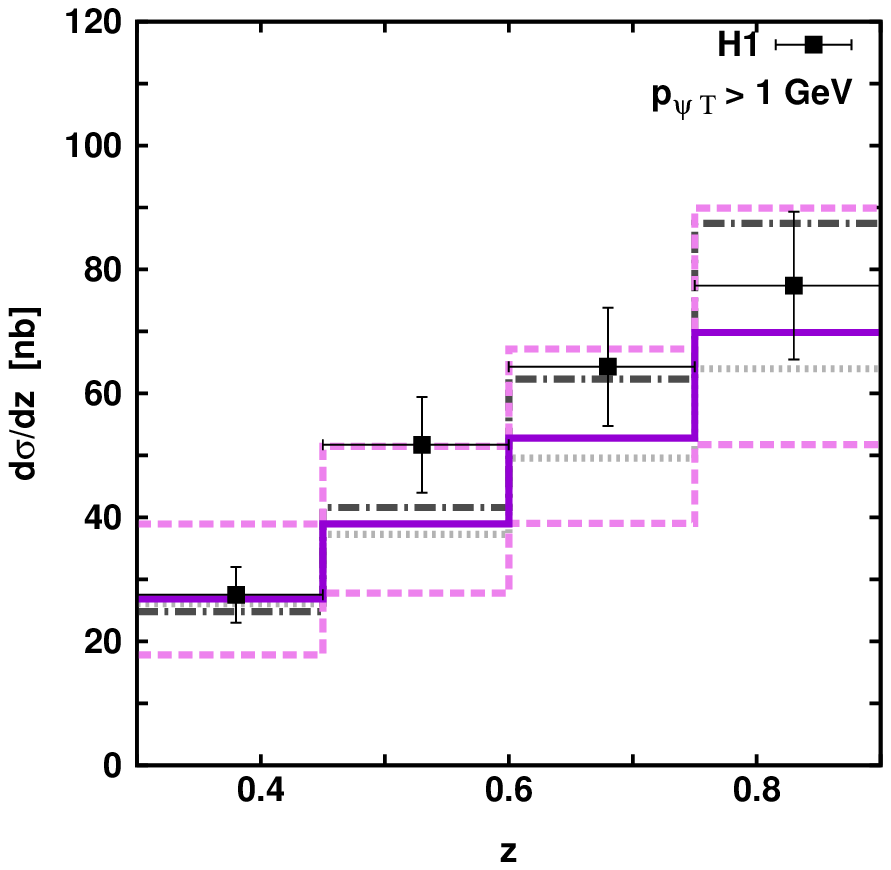, width = 8.1cm}
\epsfig{figure=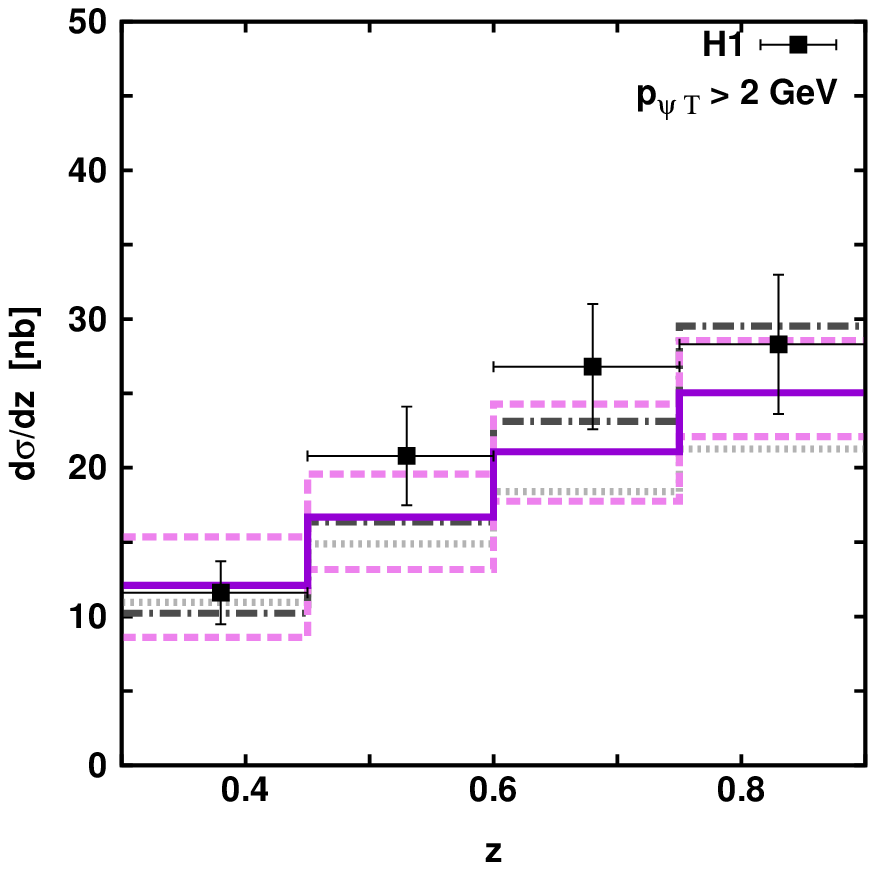, width = 8.1cm}
\epsfig{figure=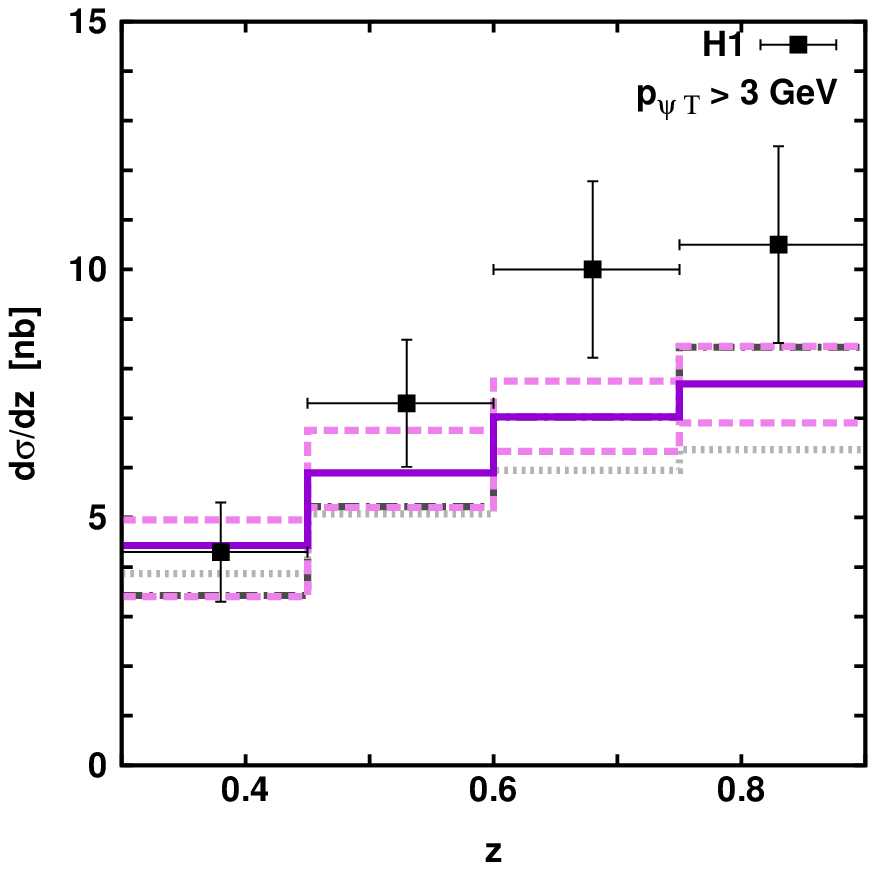, width = 8.1cm}
\epsfig{figure=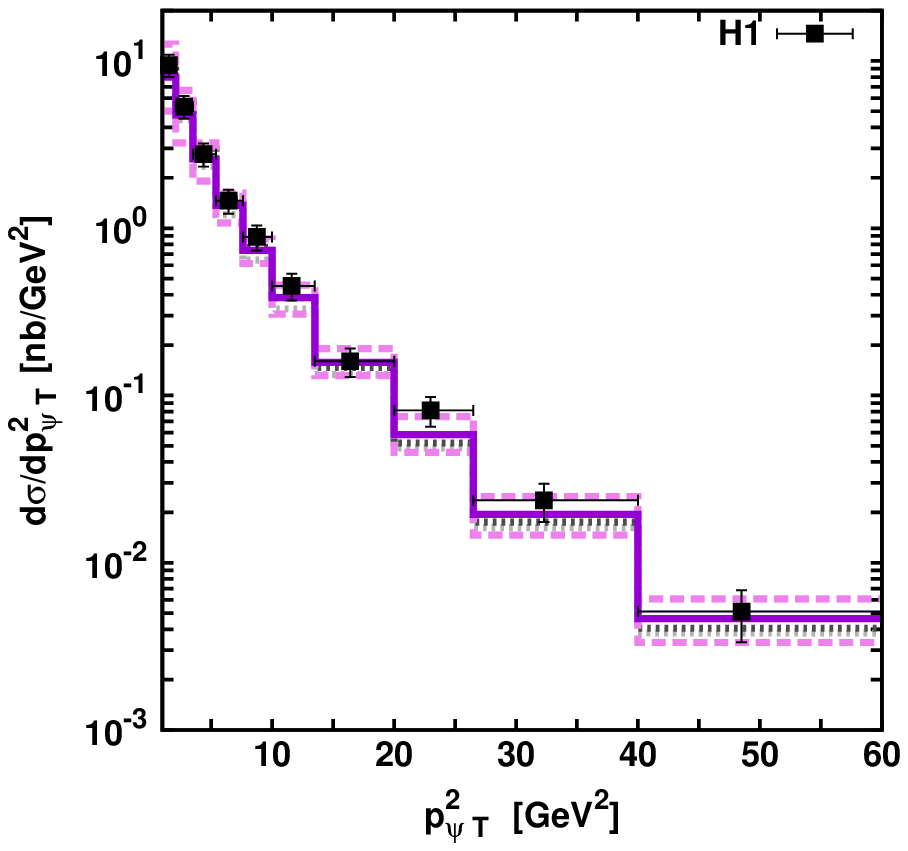, width = 8.1cm}
\epsfig{figure=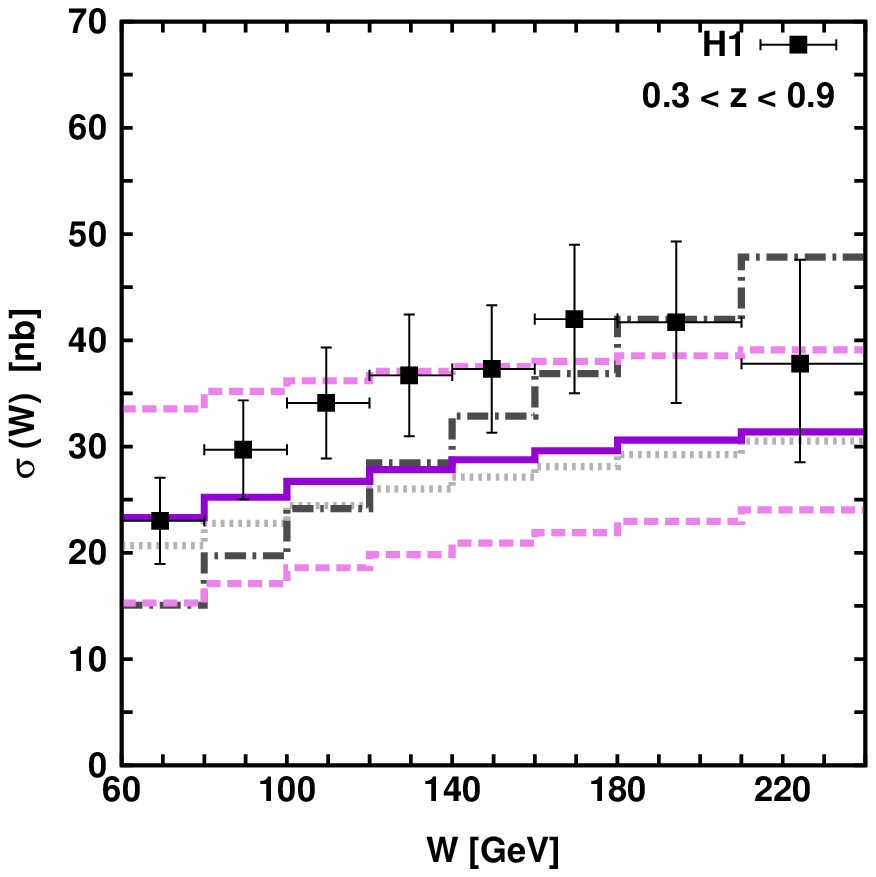, width = 8.1cm}
\epsfig{figure=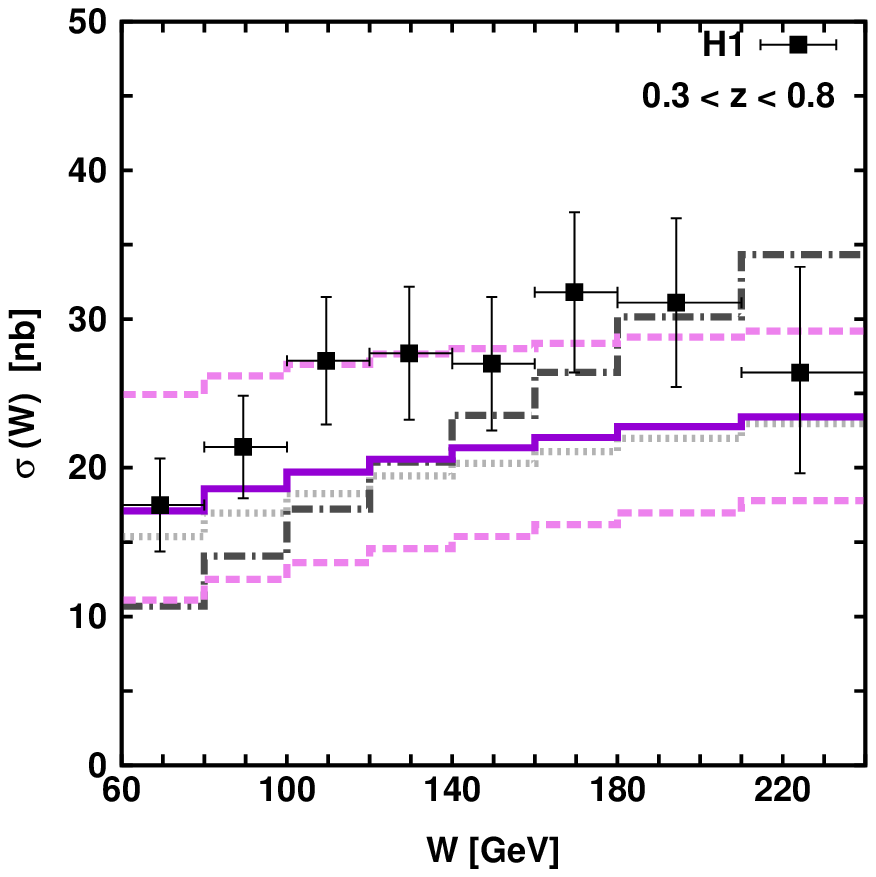, width = 8.1cm}
\caption{The total and differential cross sections of $J/\psi$ mesons calculated
in the kinematical region defined by $0.3 < z < 0.9$, 
${\mathbf p}_{\psi\,T}^2 > 1$~GeV$^2$ and $60 < W < 240$~GeV.
Notation of all histograms is the same as in Fig.~1.
The experimental data are from H1~\cite{28}.}
\end{center}
\label{fig4}
\end{figure}

\newpage

\begin{figure}
\begin{center}
\epsfig{figure=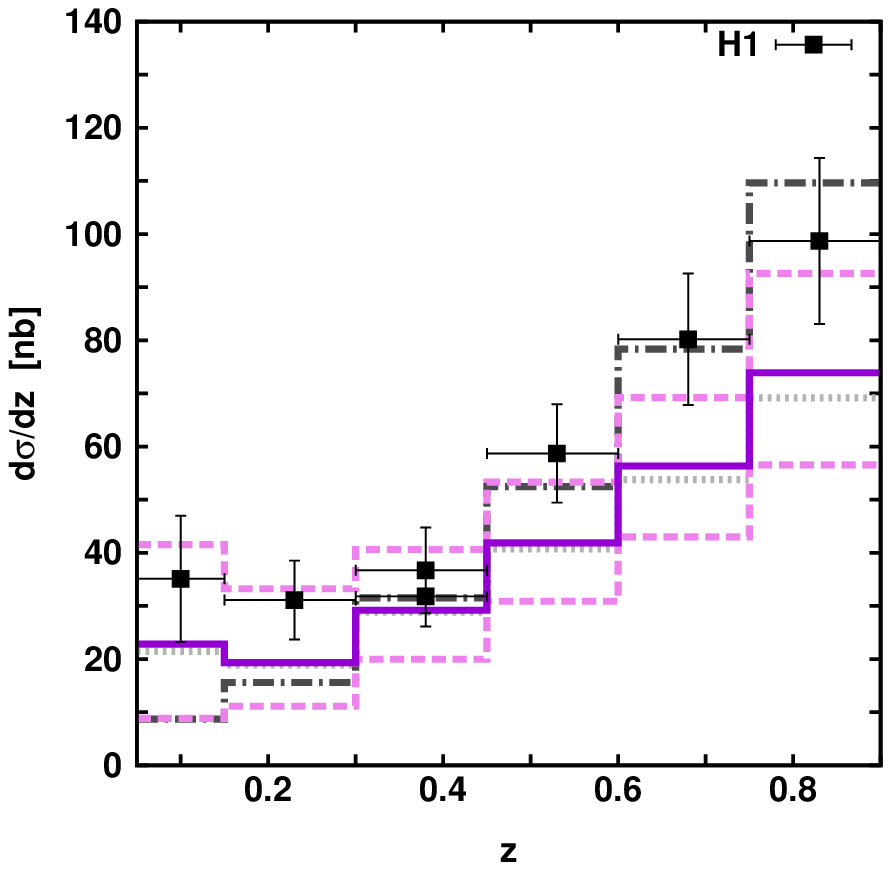, width = 8.1cm}
\epsfig{figure=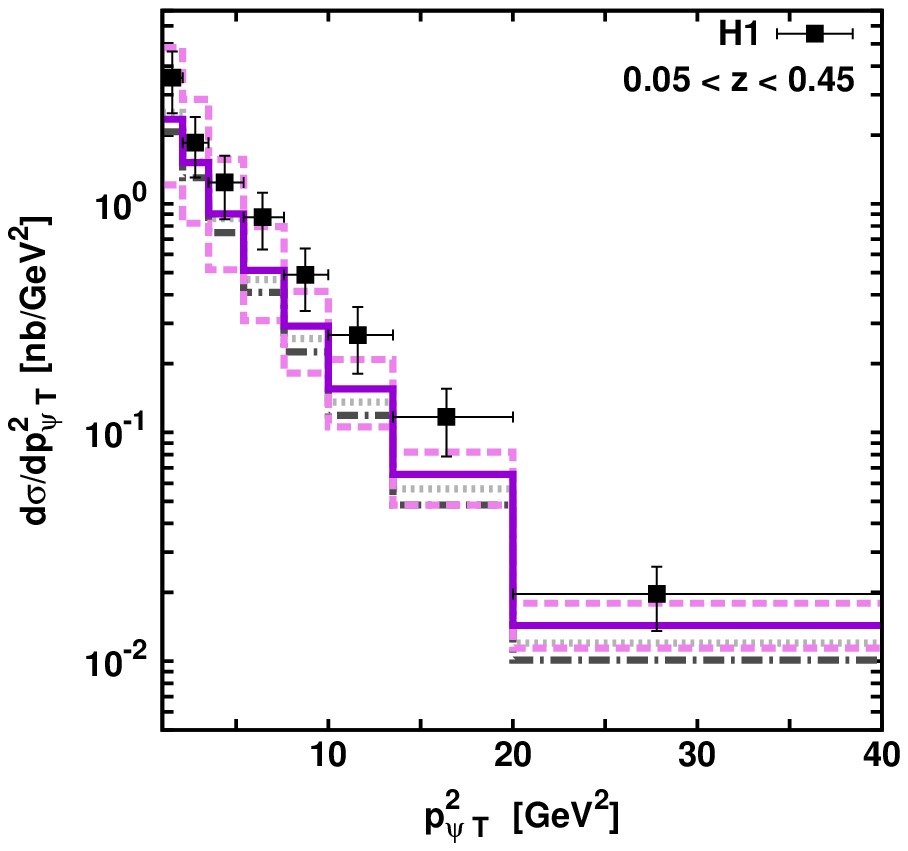, width = 8.1cm}
\epsfig{figure=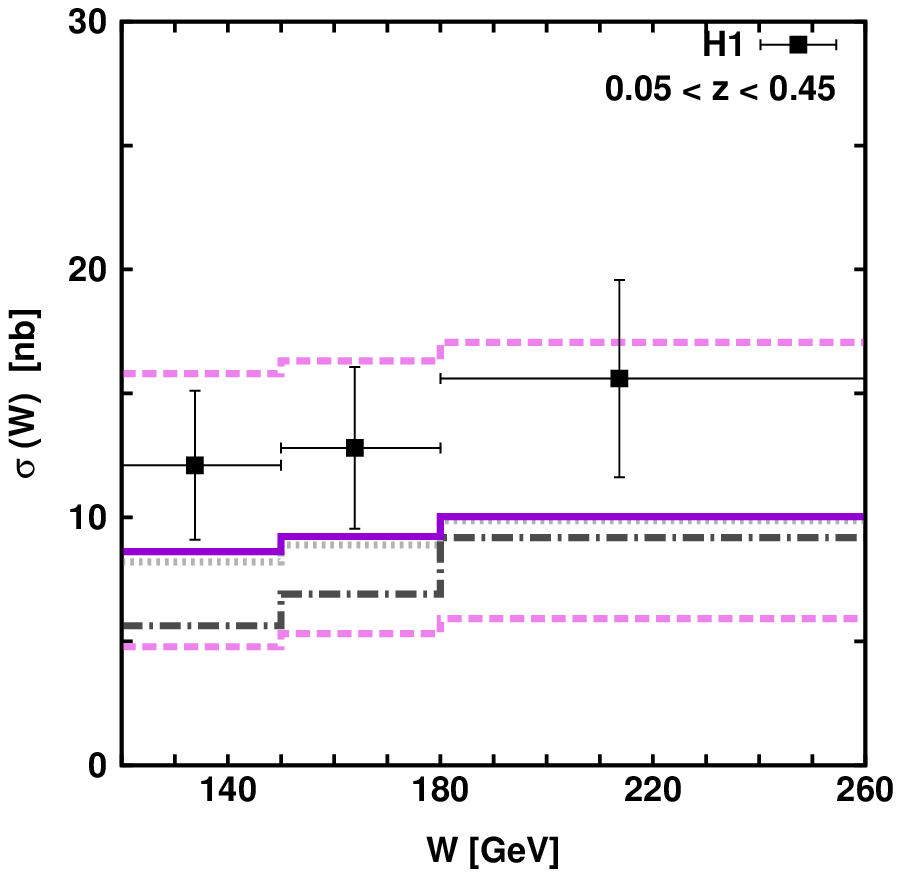, width = 8.1cm}
\caption{The total and differential cross sections of $J/\psi$ mesons calculated
in the kinematical region defined by $0.05 < z < 0.9$, 
${\mathbf p}_{\psi\,T}^2 > 1$~GeV$^2$ and $120 < W < 260$~GeV.
Notation of all histograms is the same as in Fig.~1.
The experimental data are from H1~\cite{28}.}
\end{center}
\label{fig5}
\end{figure}

\newpage

\begin{figure}
\begin{center}
\epsfig{figure=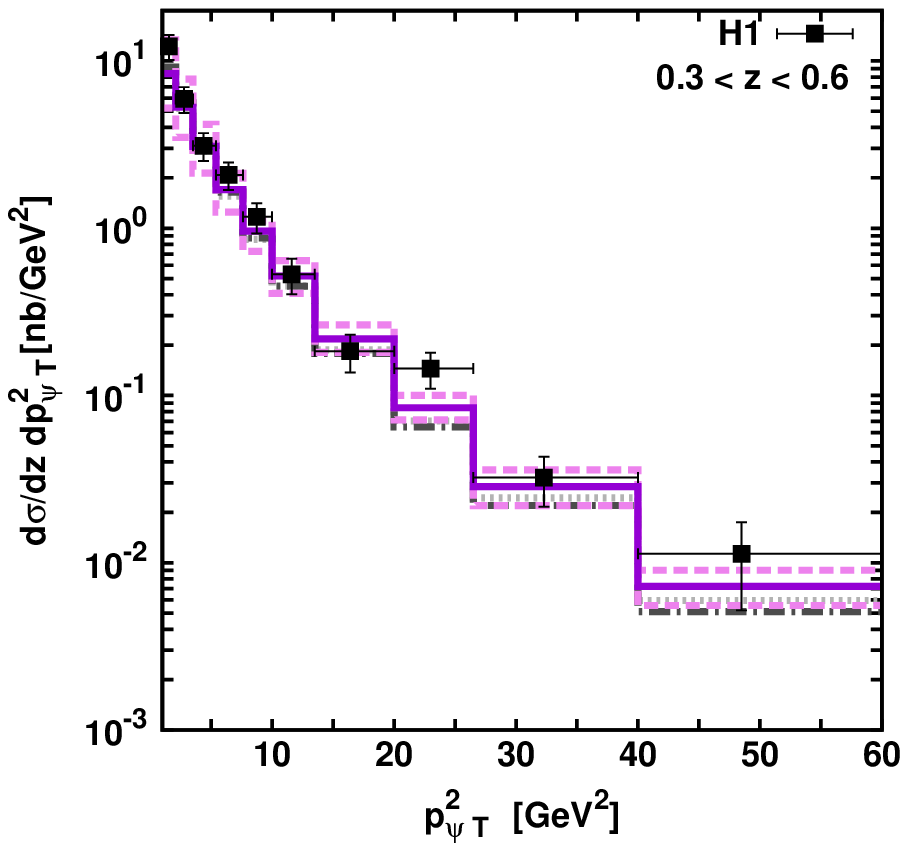, width = 8.1cm}
\epsfig{figure=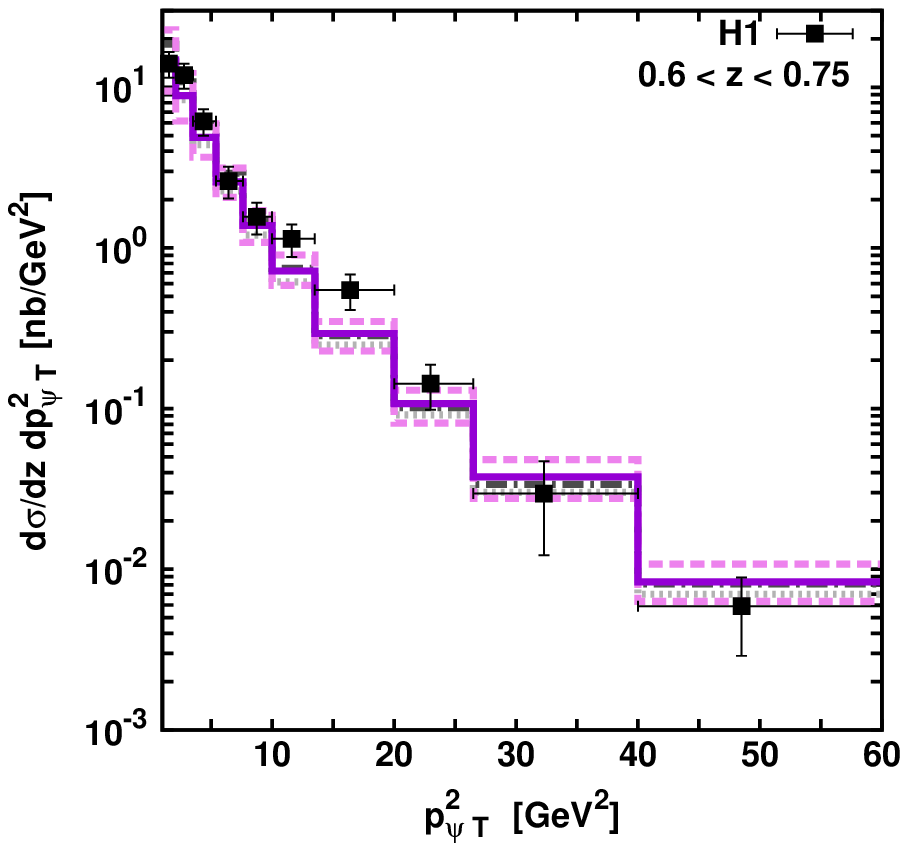, width = 8.1cm}
\epsfig{figure=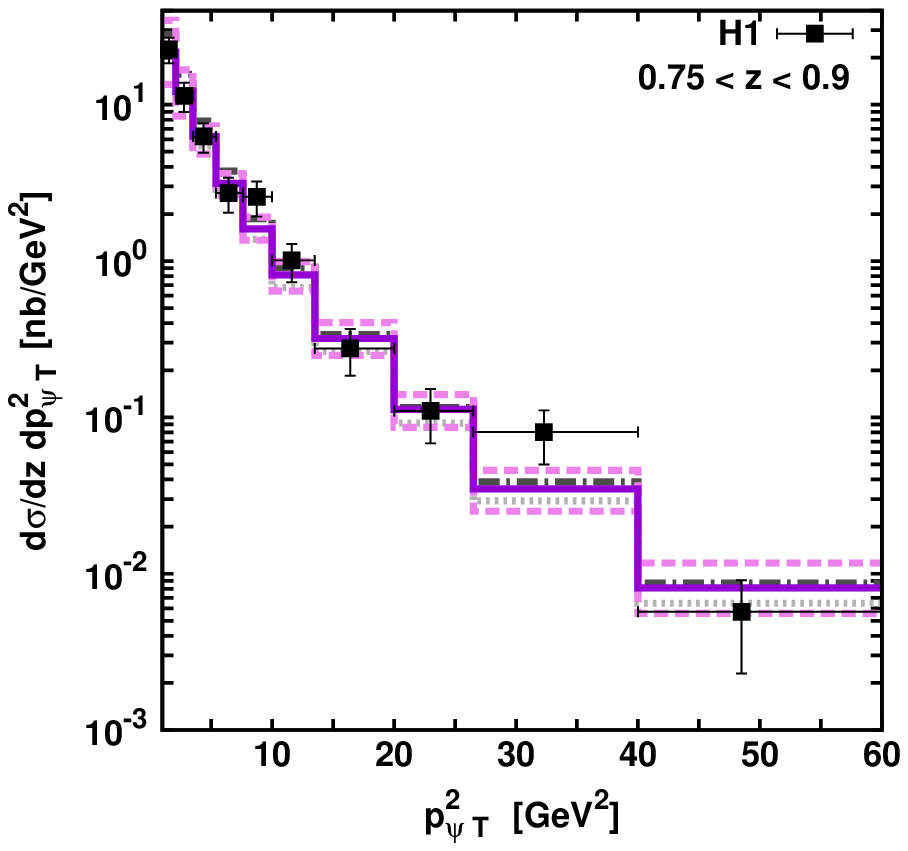, width = 8.1cm}
\caption{The double differential cross sections of $J/\psi$ mesons
in bins of $z$ and ${\mathbf p}_{\psi \,T}^2$ calculated 
at $60 < W < 240$~GeV.
Notation of all histograms is the same as in Fig.~1.
The experimental data are from H1~\cite{28}.}
\end{center}
\label{fig6}
\end{figure}

\newpage

\begin{figure}
\begin{center}
\epsfig{figure=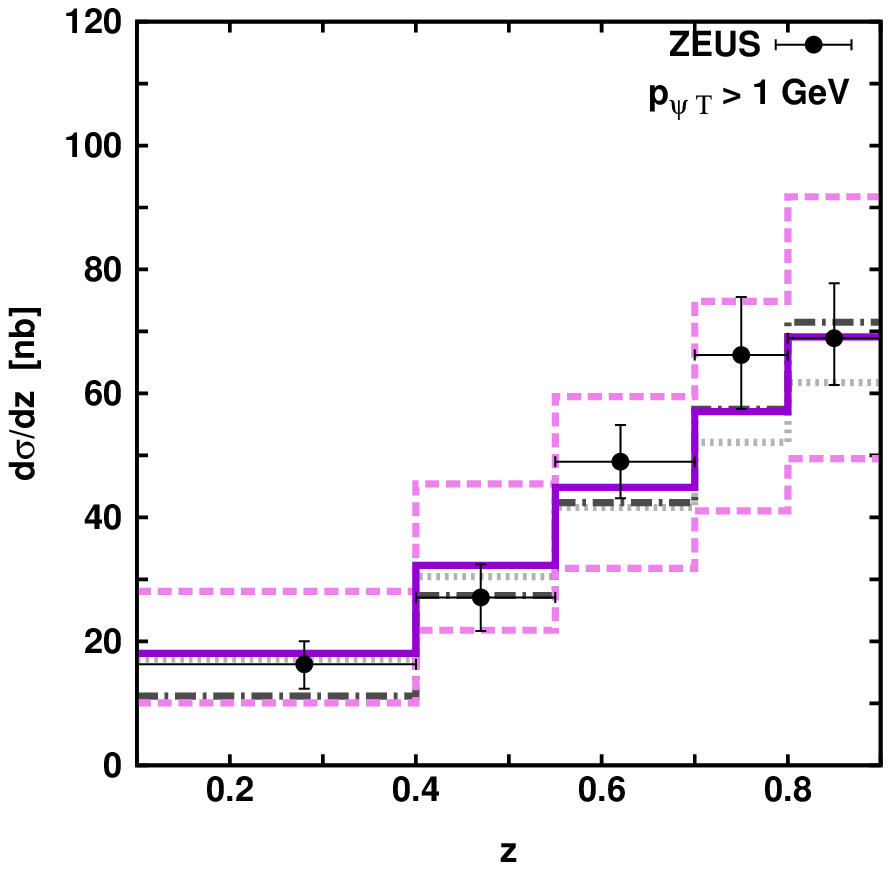, width = 8.1cm}
\epsfig{figure=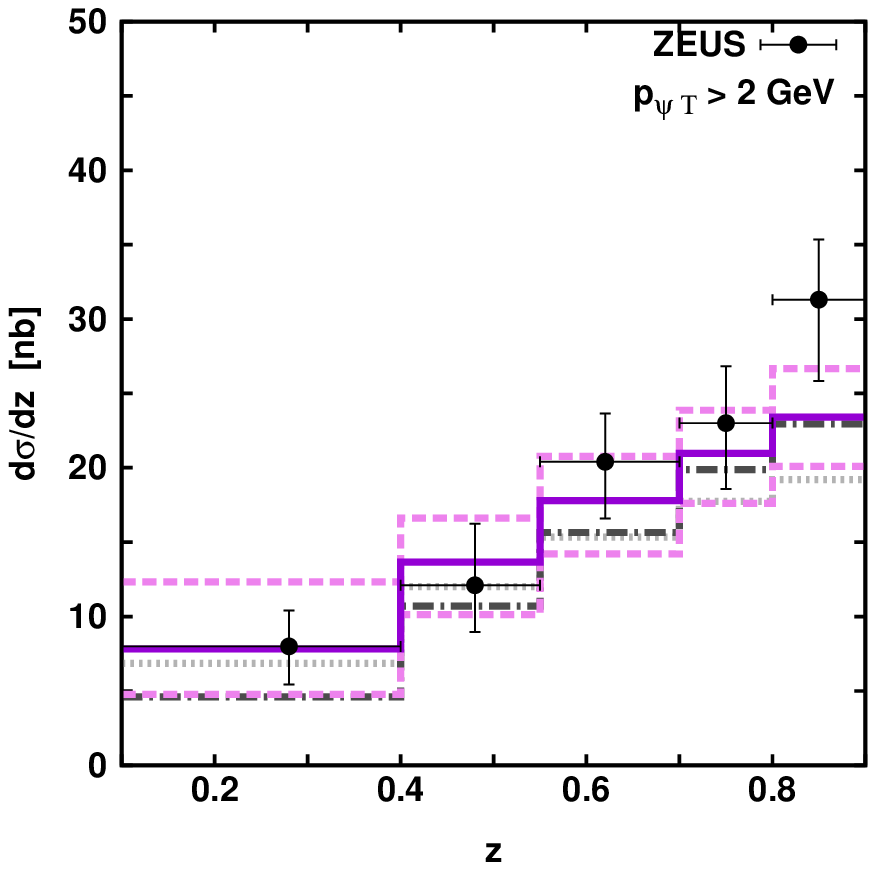, width = 8.1cm}
\epsfig{figure=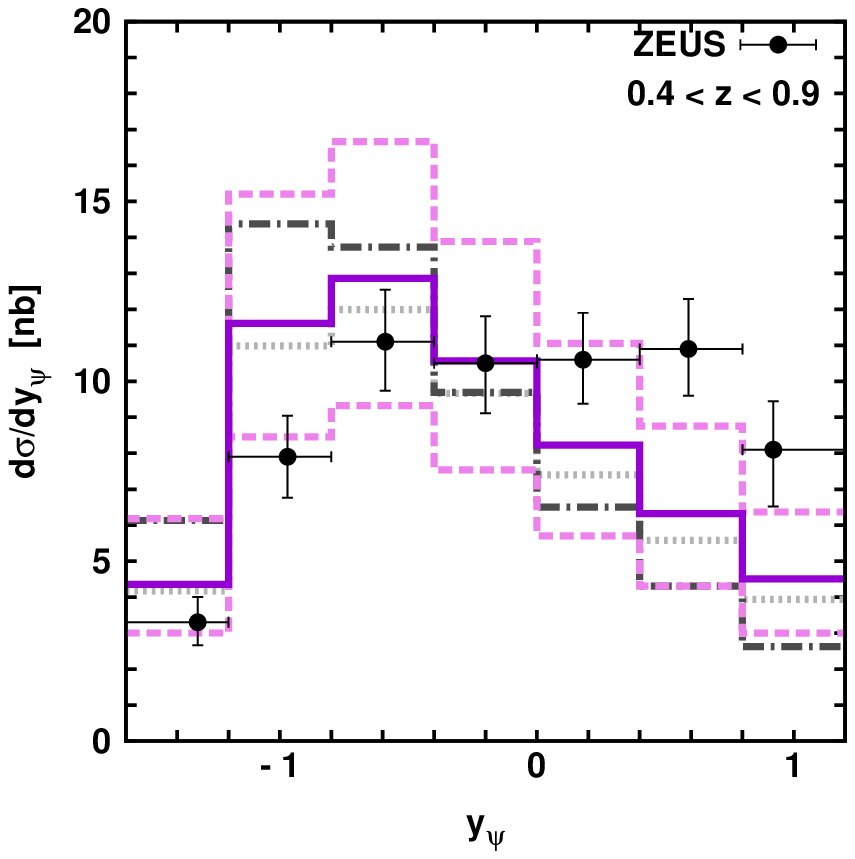, width = 8.1cm}
\epsfig{figure=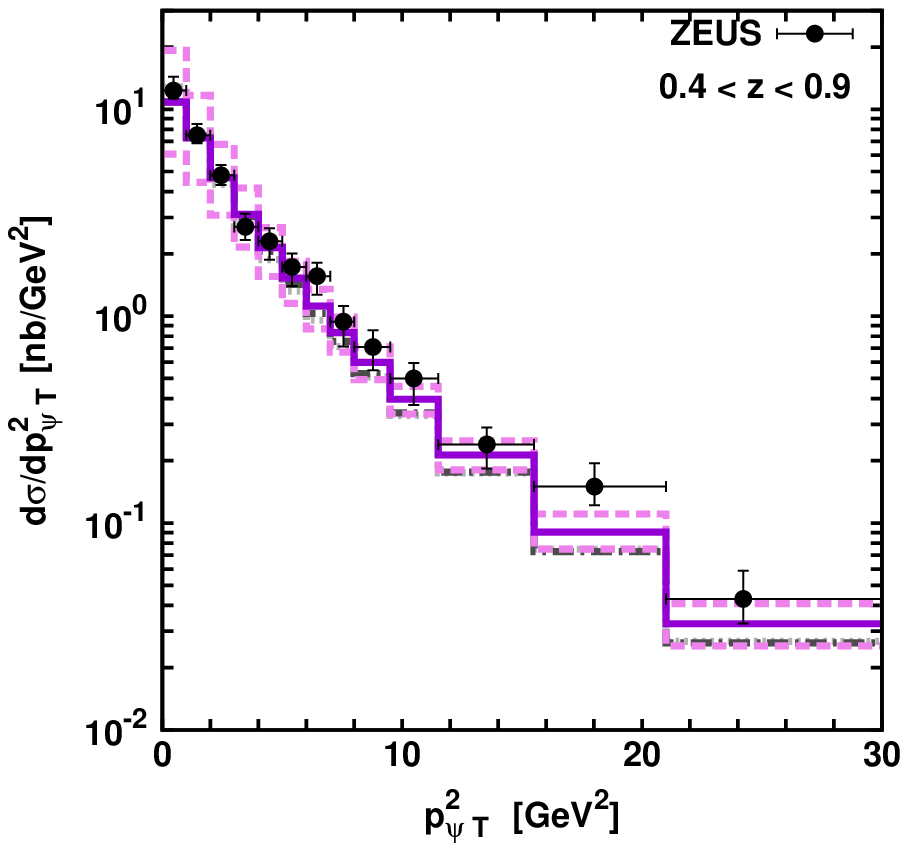, width = 8.1cm}
\epsfig{figure=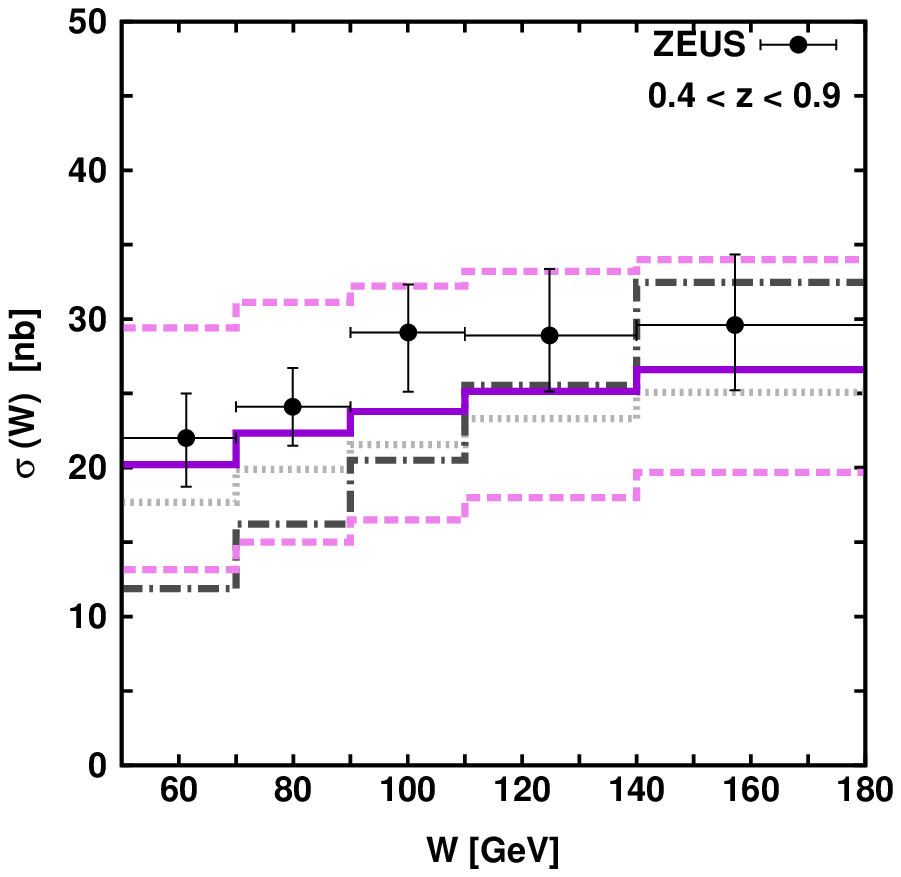, width = 8.1cm}
\caption{The total and differential cross sections of $J/\psi$ mesons calculated
in the kinematical region defined by $0.1 < z < 0.9$, 
${\mathbf p}_{\psi\,T}^2 > 1$~GeV$^2$ and $50 < W < 180$~GeV.
Notation of all histograms is the same as in Fig.~1.
The experimental data are from ZEUS~\cite{29}.}
\end{center}
\label{fig7}
\end{figure}

\newpage

\begin{figure}
\begin{center}
\epsfig{figure=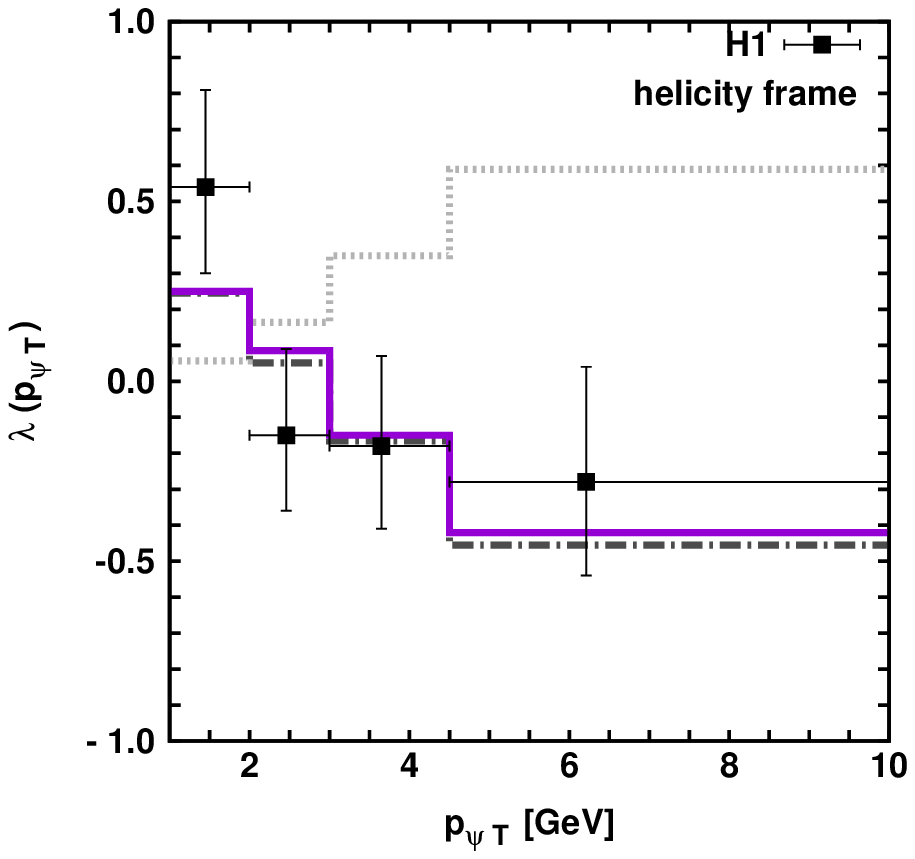, width = 8.1cm}
\epsfig{figure=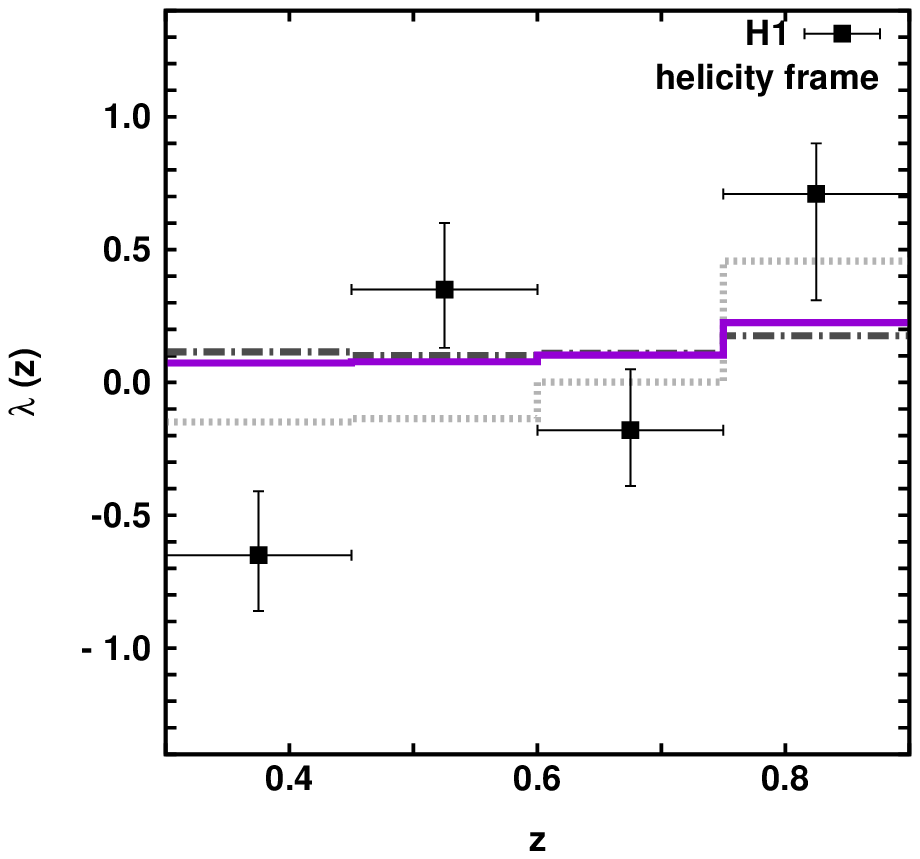, width = 8.1cm}
\epsfig{figure=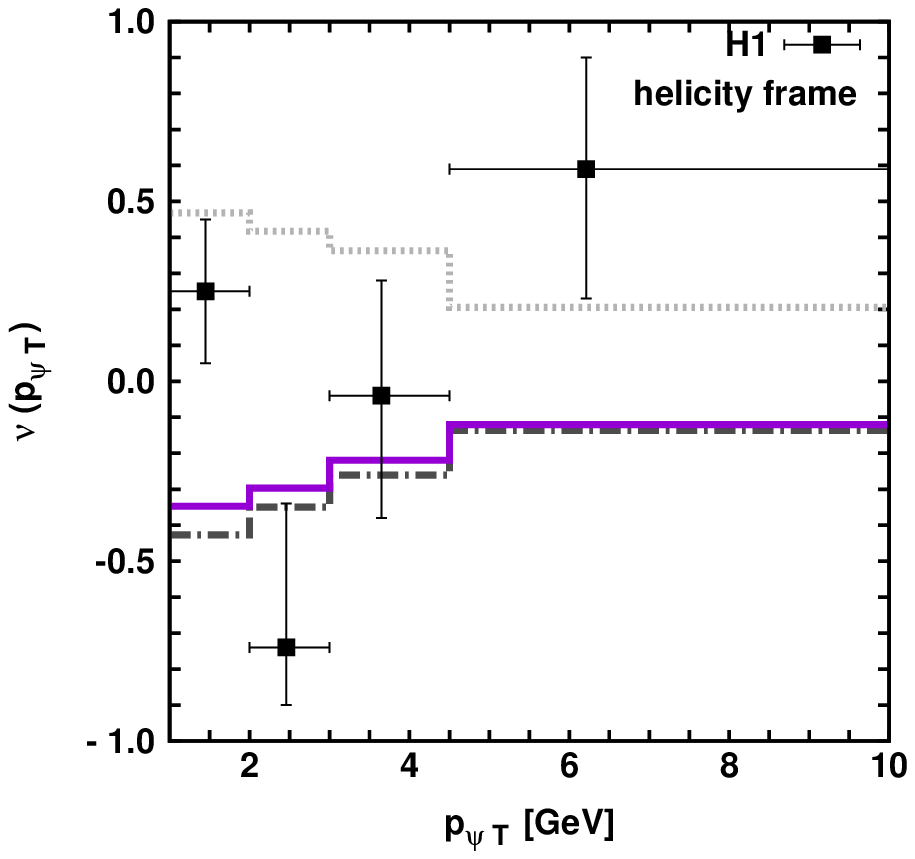, width = 8.1cm}
\epsfig{figure=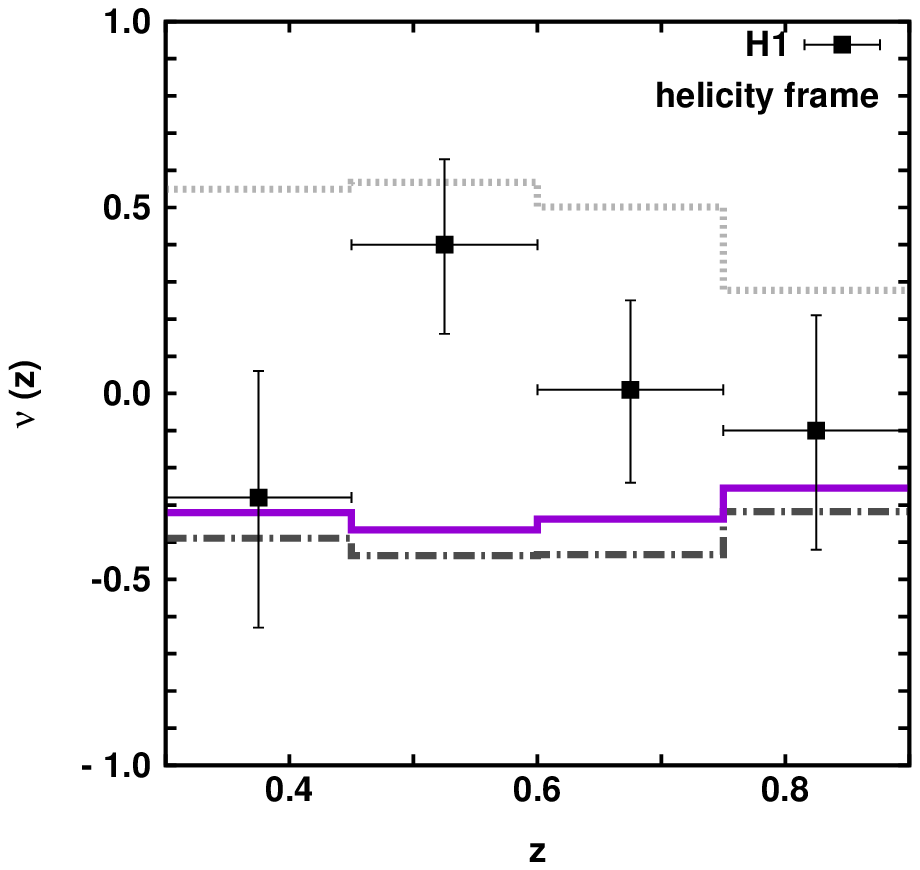, width = 8.1cm}
\caption{Polarization parameters $\lambda$ and $\nu$ as a 
functions of the $J/\psi$ transverse momentum and 
elasticity $z$ calculated in the helicity frame at
$0.3 < z < 0.9$, ${\mathbf p}_{\psi\,T}^2 > 1$~GeV$^2$ and $60 < W < 240$~GeV.
The solid and dash-dotted histograms correspond to the results obtained 
using the CCFM A0 and KMR gluon densities. The dotted histograms
represent the LO CS predictions. The experimental data are from H1~\cite{27}.}
\end{center}
\label{fig8}
\end{figure}

\begin{figure}
\begin{center}
\epsfig{figure=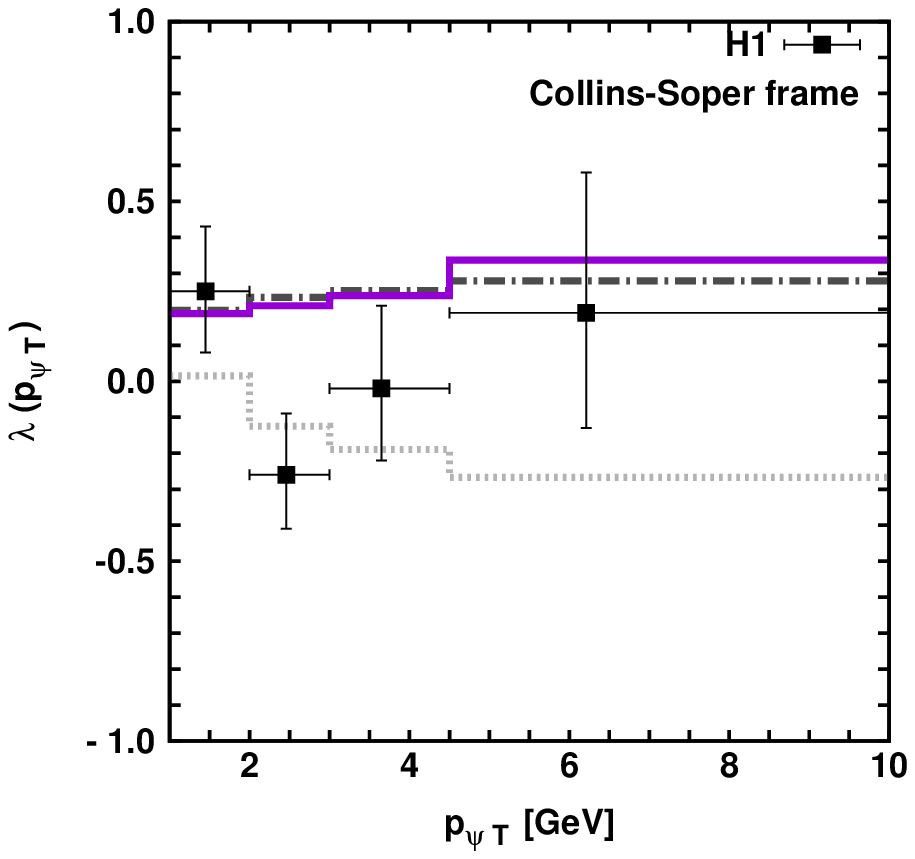, width = 8.1cm}
\epsfig{figure=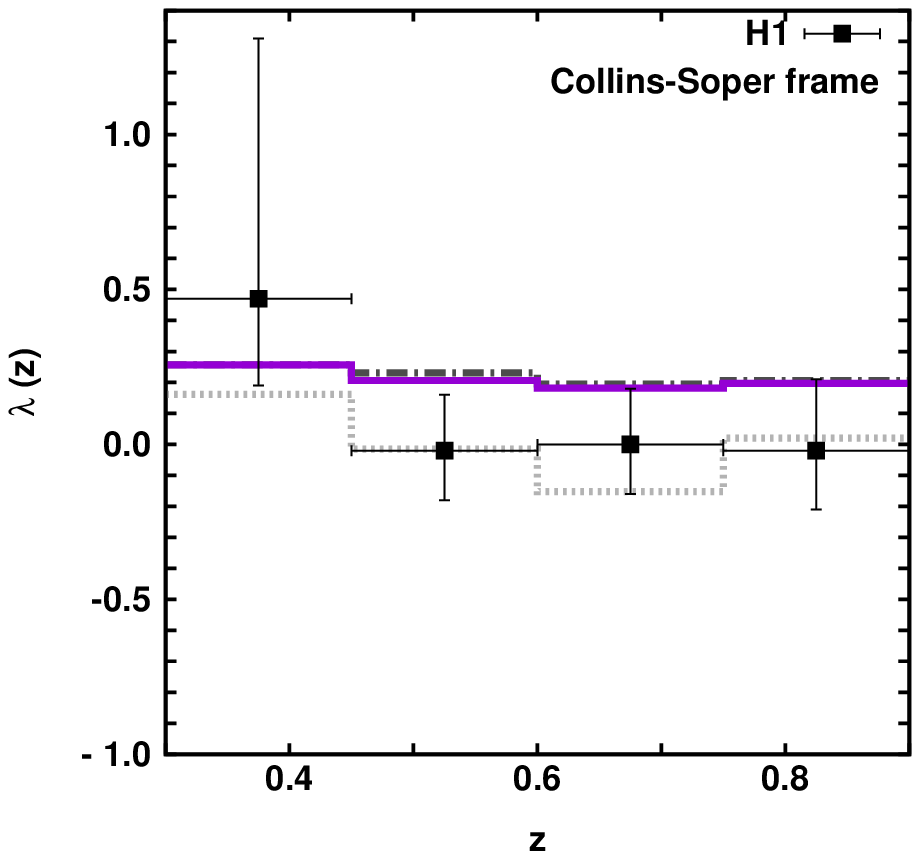, width = 8.1cm}
\epsfig{figure=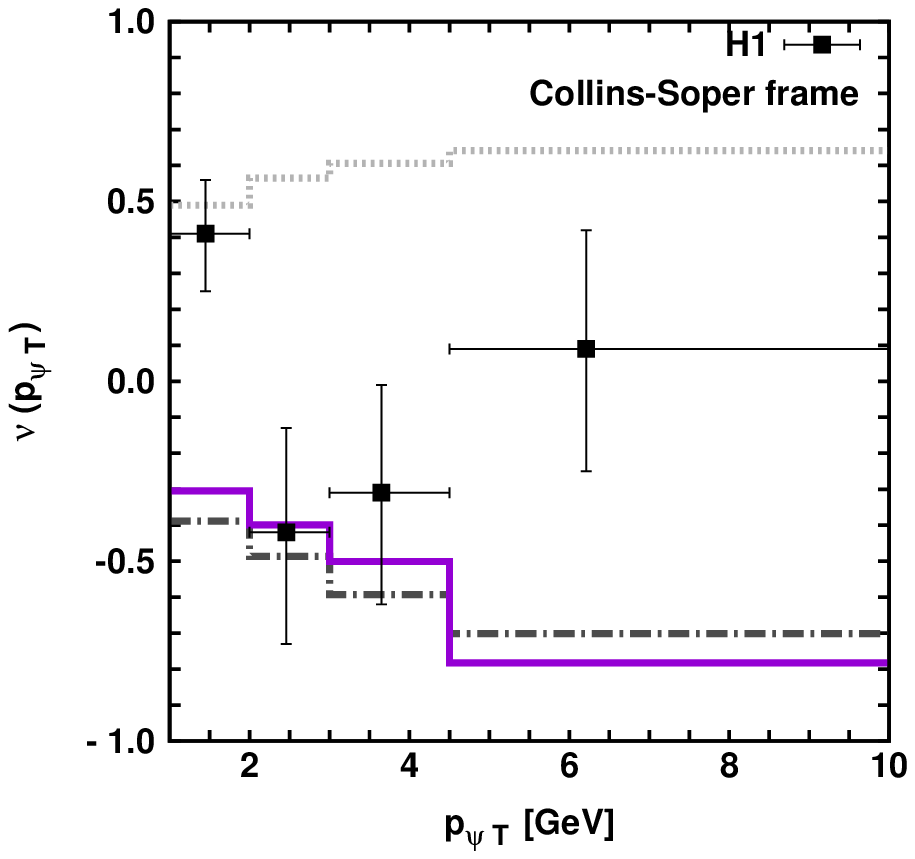, width = 8.1cm}
\epsfig{figure=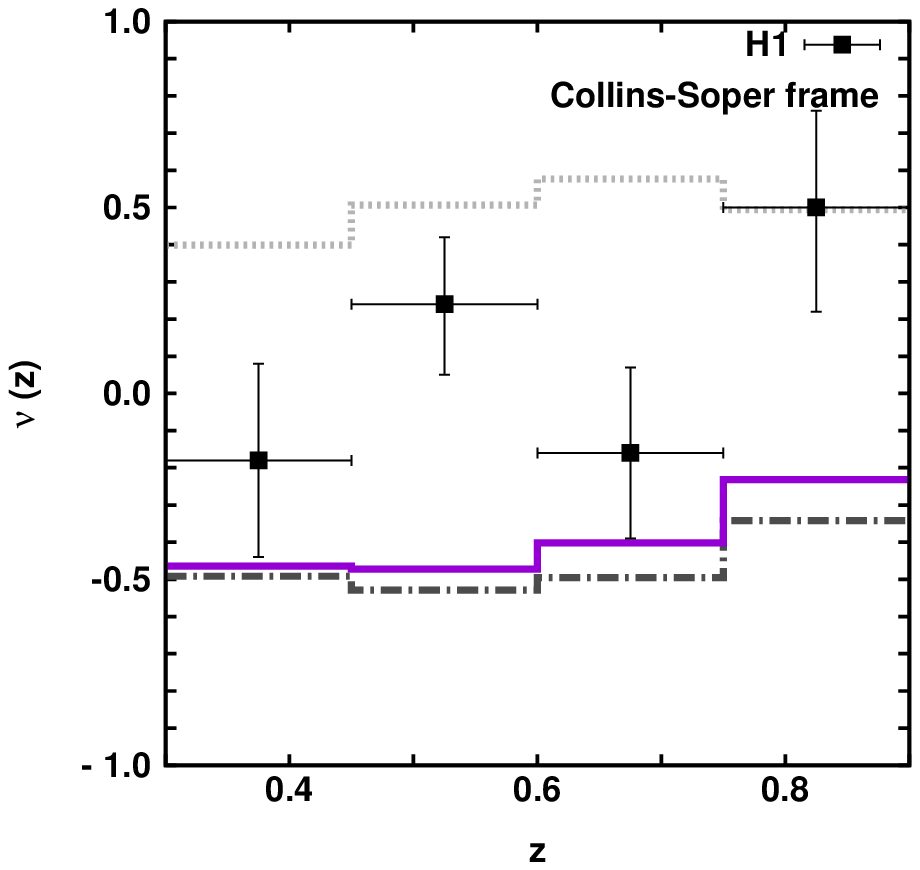, width = 8.1cm}
\caption{Polarization parameters $\lambda$ and $\nu$ as a 
functions of the $J/\psi$ transverse momentum and 
elasticity $z$ calculated in the Collins-Soper frame at
$0.3 < z < 0.9$, ${\mathbf p}_{\psi\,T}^2 > 1$~GeV$^2$ and $60 < W < 240$~GeV.
Notation of all histograms is the same as in Fig.~8.
The experimental data are from H1~\cite{27}.}
\end{center}
\label{fig9}
\end{figure}

\begin{figure}
\begin{center}
\epsfig{figure=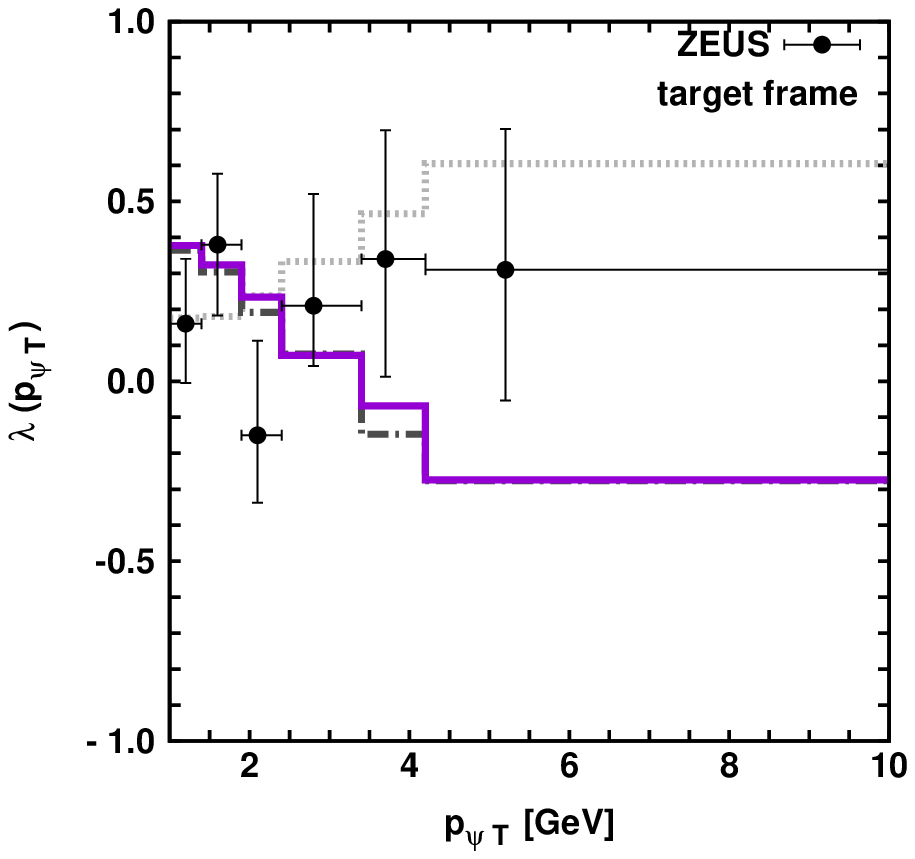, width = 8.1cm}
\epsfig{figure=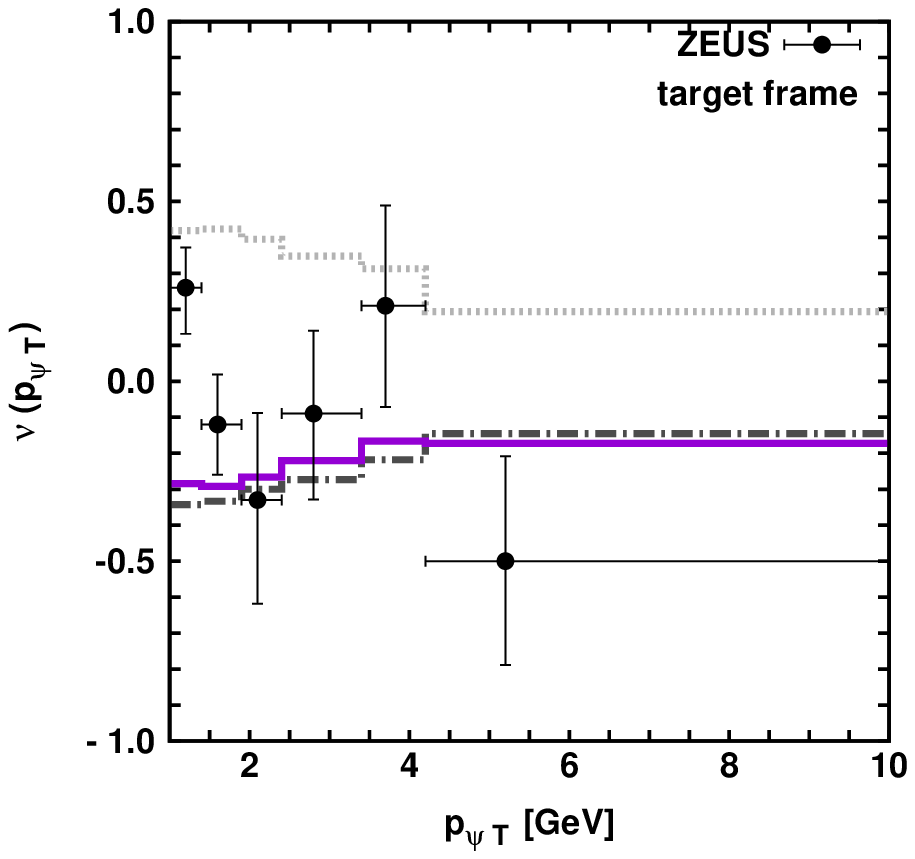, width = 8.1cm}
\caption{Polarization parameters $\lambda$ and $\nu$ as a 
functions of the $J/\psi$ transverse momentum 
calculated in the target frame at
$0.4 < z < 1.0$ and $50 < W < 180$~GeV.
Notation of all histograms is the same as in Fig.~8.
The experimental data are from ZEUS~\cite{26}.}
\end{center}
\label{fig10}
\end{figure}

\begin{figure}
\begin{center}
\epsfig{figure=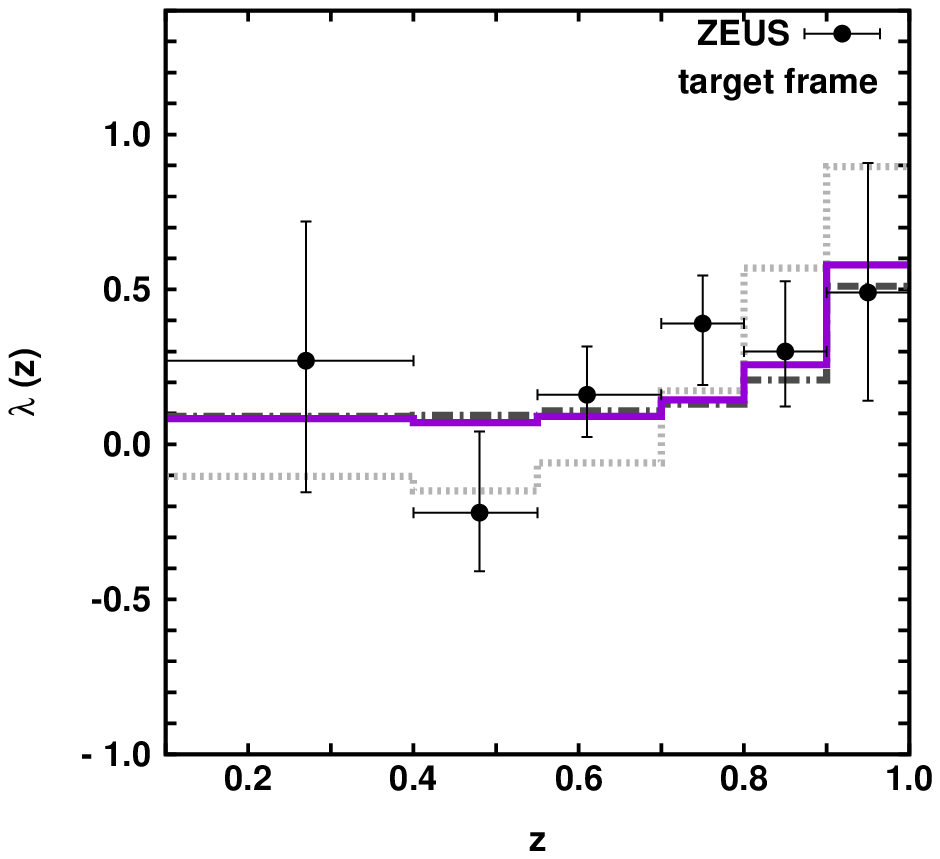, width = 8.1cm}
\epsfig{figure=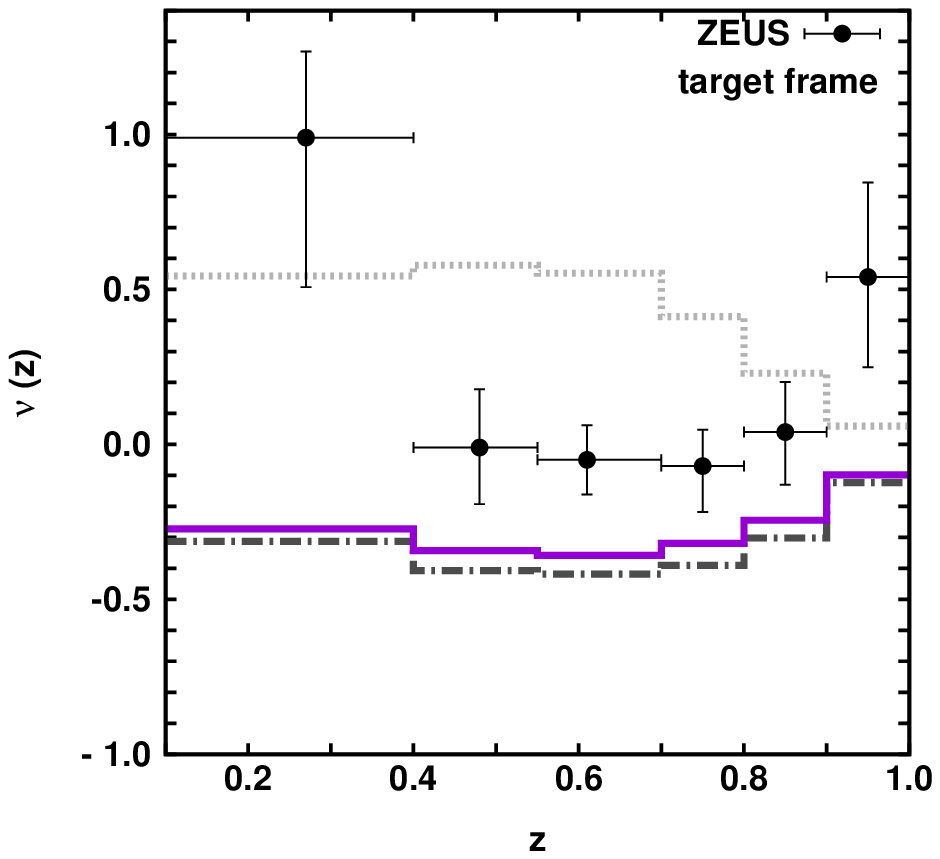, width = 8.1cm}
\caption{Polarization parameters $\lambda$ and $\nu$ as a 
functions of the elasticity $z$
calculated in the target frame at
$0.1 < z < 1.0$, ${\mathbf p}_{\psi\,T} > 1$~GeV and $50 < W < 180$~GeV.
Notation of all histograms is the same as in Fig.~8.
The experimental data are from ZEUS~\cite{26}.}
\end{center}
\label{fig11}
\end{figure}

\begin{figure}
\begin{center}
\epsfig{figure=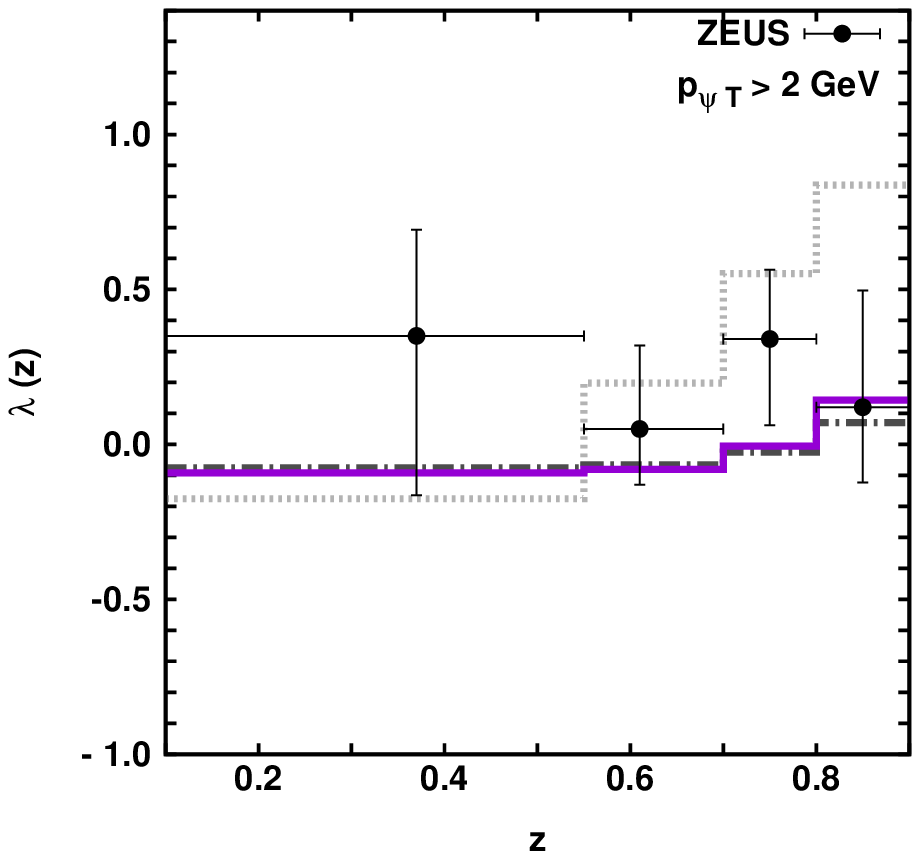, width = 8.1cm}
\epsfig{figure=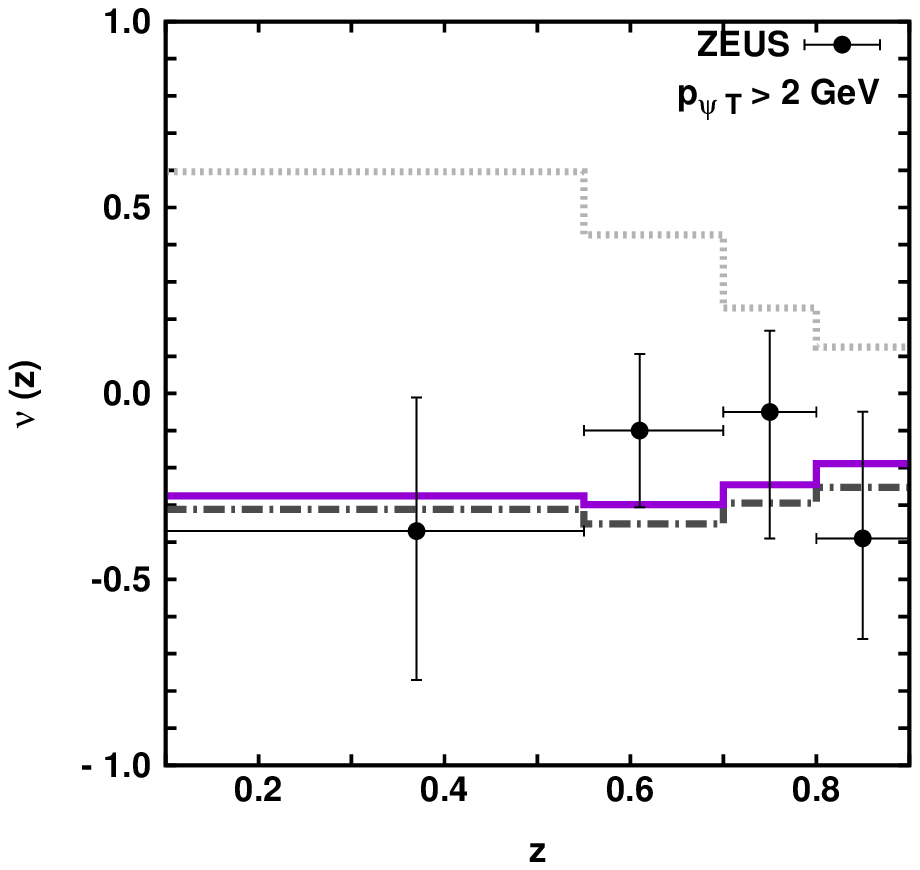, width = 8.1cm}
\epsfig{figure=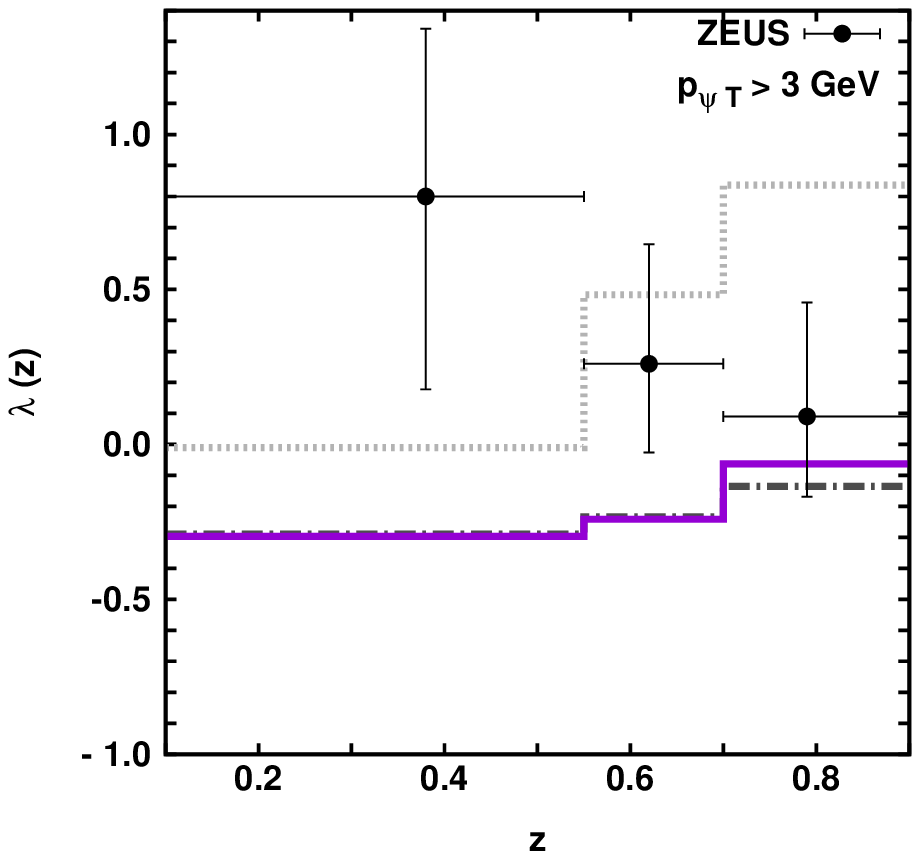, width = 8.1cm}
\epsfig{figure=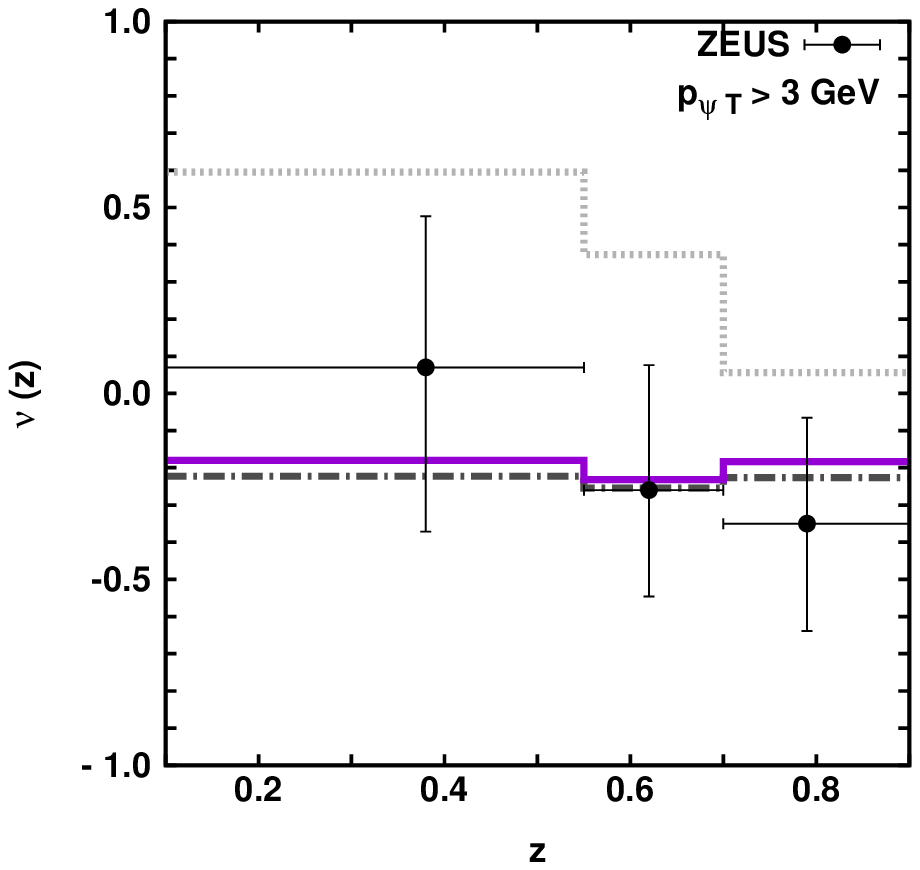, width = 8.1cm}
\caption{Polarization parameters $\lambda$ and $\nu$ as a 
functions of the elasticity $z$
calculated in the target frame at
$0.1 < z < 0.9$ and $50 < W < 180$~GeV.
Notation of all histograms is the same as in Fig.~8.
The experimental data are from ZEUS~\cite{26}.}
\end{center}
\label{fig12}
\end{figure}

\begin{figure}
\begin{center}
\epsfig{figure=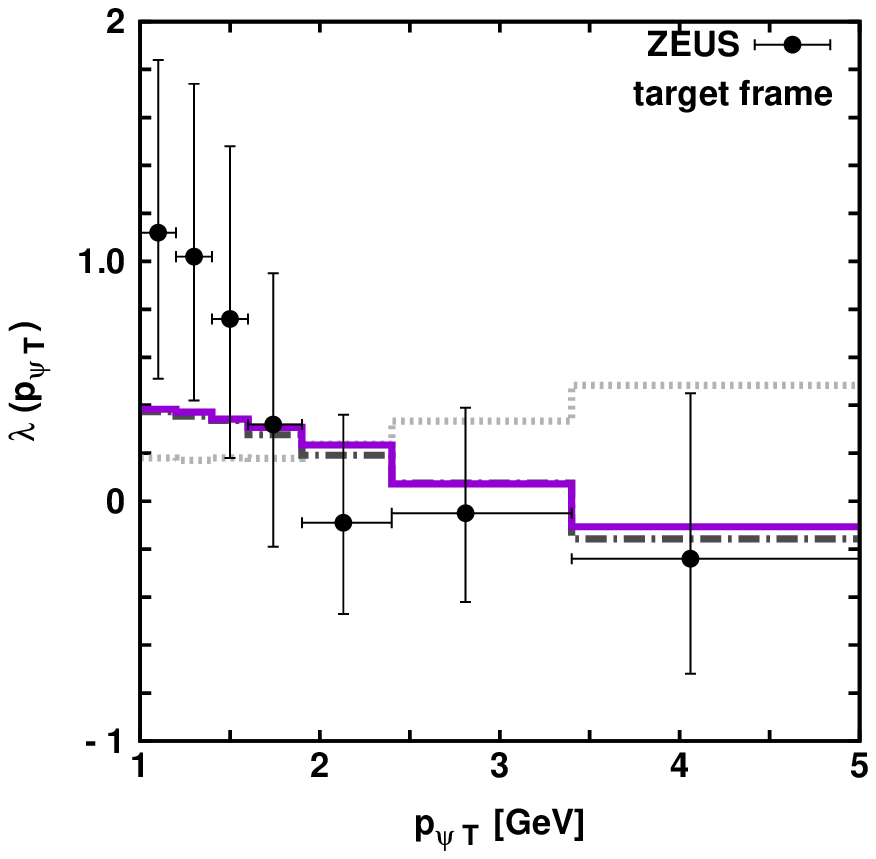, width = 8.1cm}
\epsfig{figure=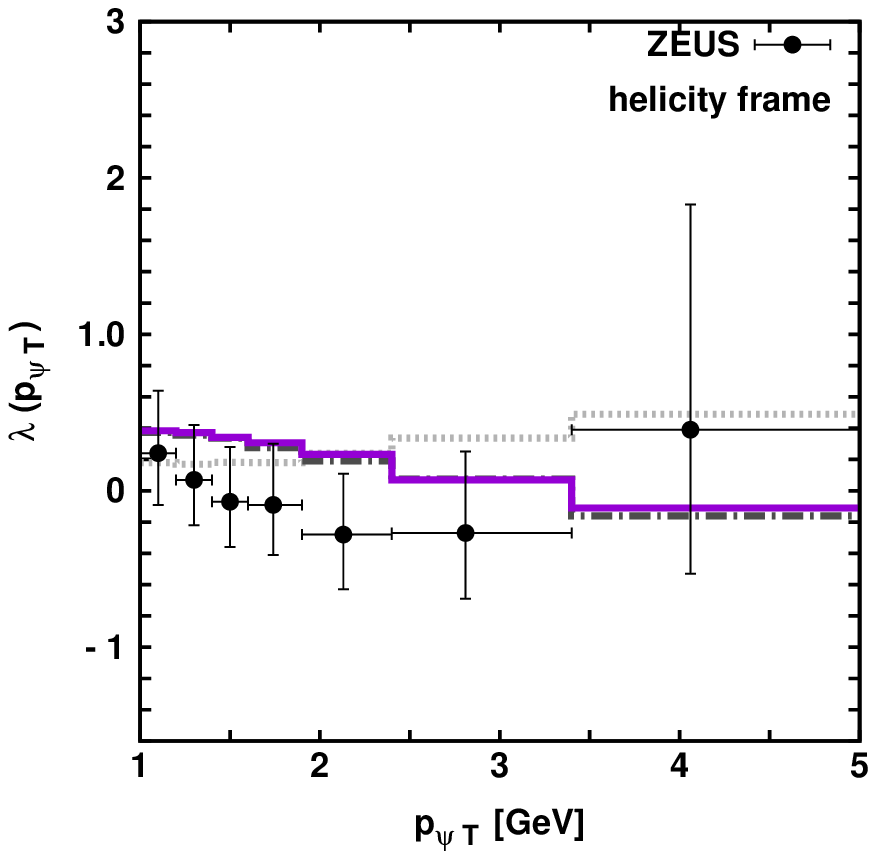, width = 8.1cm}
\caption{Polarization parameter $\lambda$ as a 
function of the $J/\psi$ transverse momentum 
calculated at $0.4 < z < 1.0$ and $50 < W < 180$~GeV.
Notation of all histograms is the same as in Fig.~8.
The experimental data are from ZEUS~\cite{29}.}
\end{center}
\label{fig13}
\end{figure}

\begin{figure}
\begin{center}
\epsfig{figure=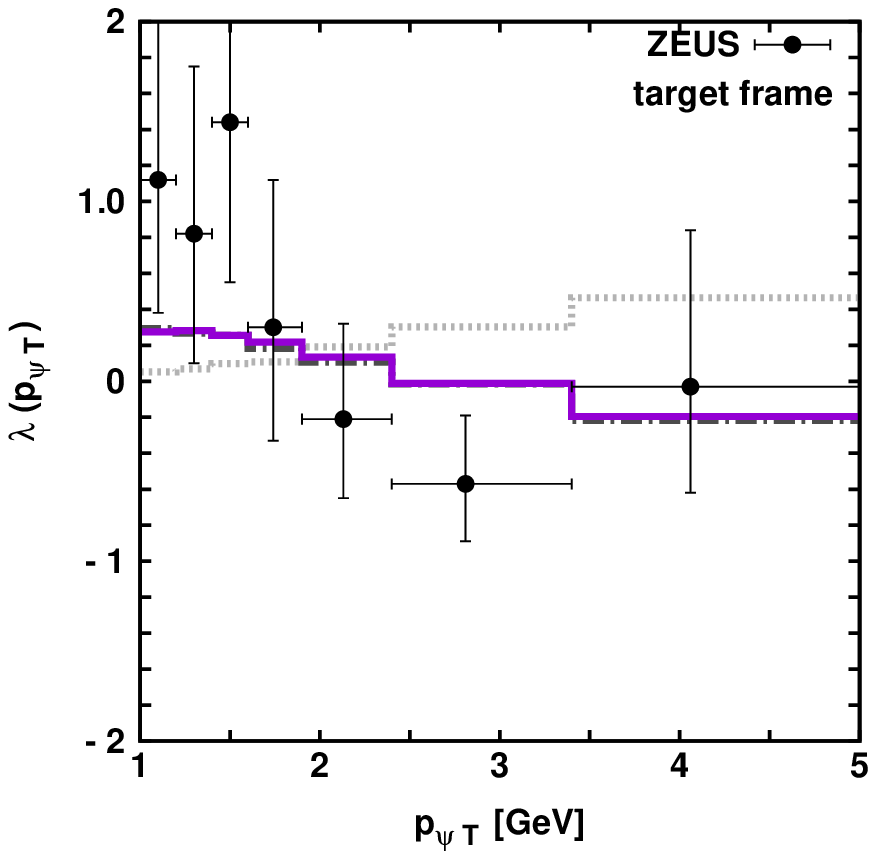, width = 8.1cm}
\epsfig{figure=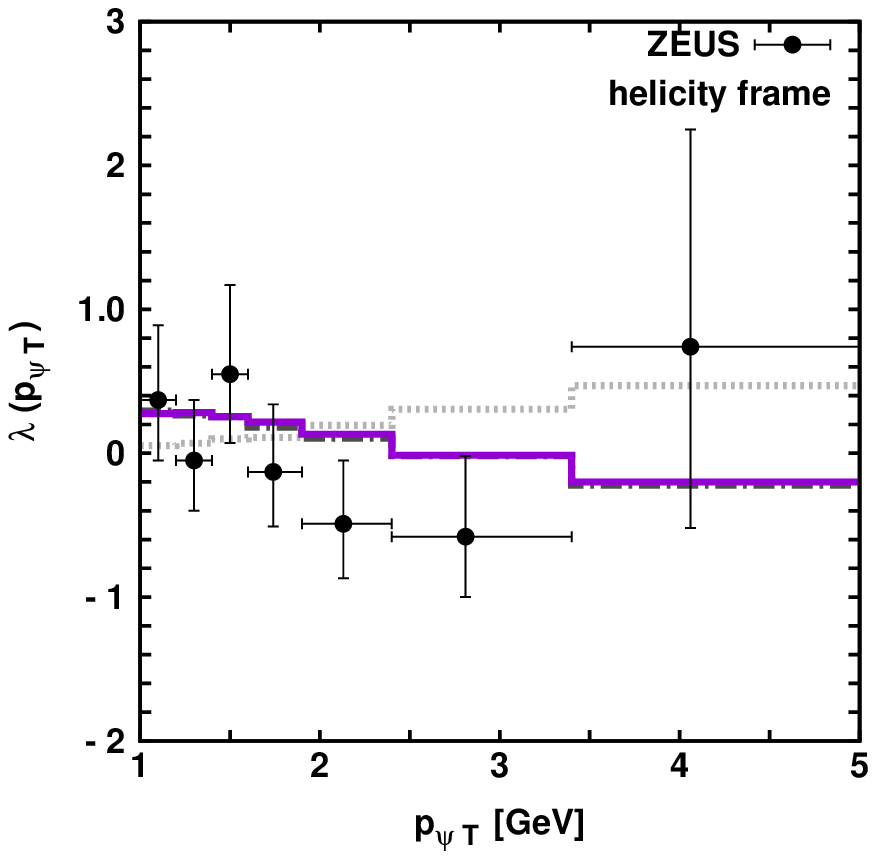, width = 8.1cm}
\caption{Polarization parameter $\lambda$ as a 
function of the $J/\psi$ transverse momentum 
calculated at $0.4 < z < 0.9$ and $50 < W < 180$~GeV.
Notation of all histograms is the same as in Fig.~8.
The experimental data are from ZEUS~\cite{29}.}
\end{center}
\label{fig14}
\end{figure}

\begin{figure}
\begin{center}
\epsfig{figure=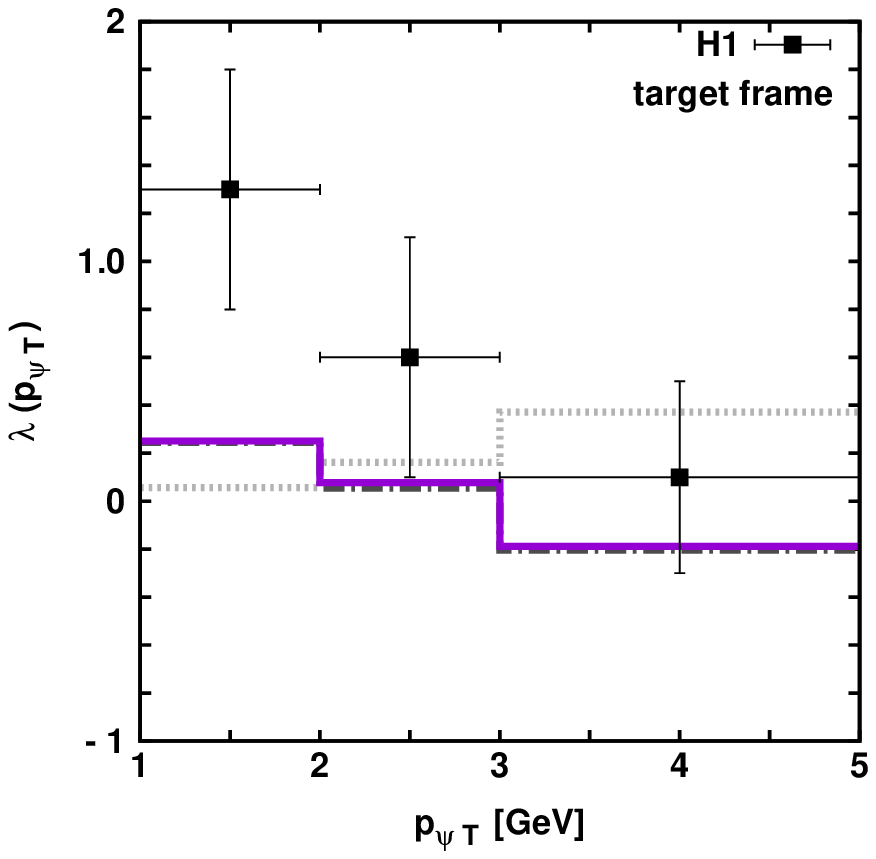, width = 8.1cm}
\epsfig{figure=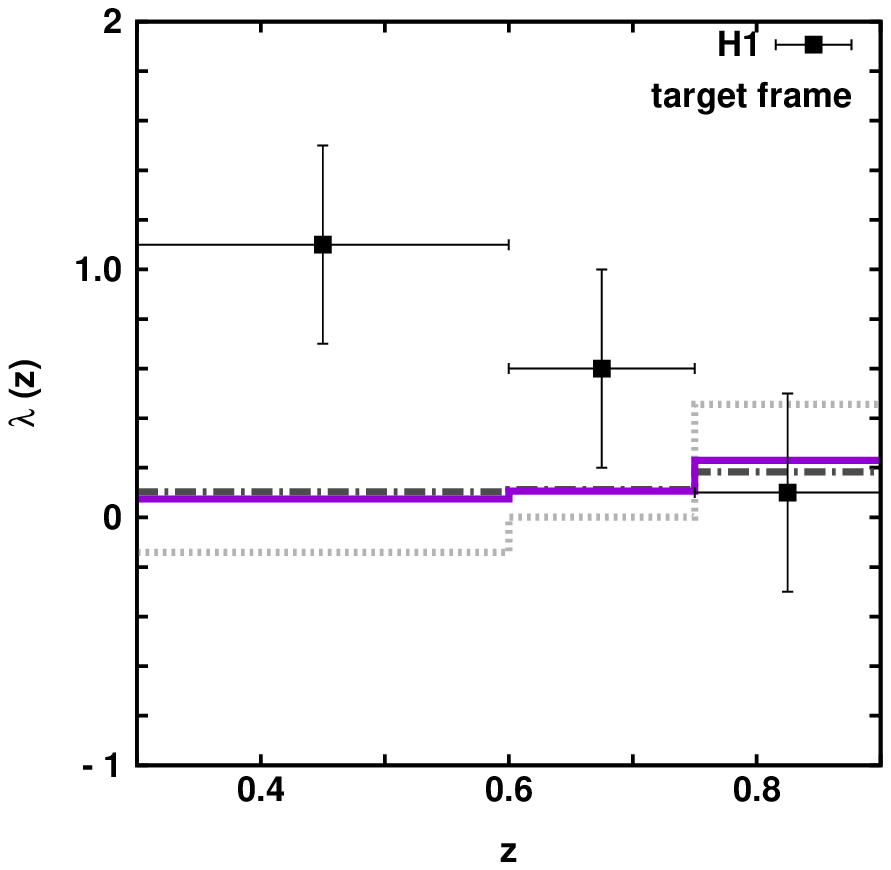, width = 8.1cm}
\epsfig{figure=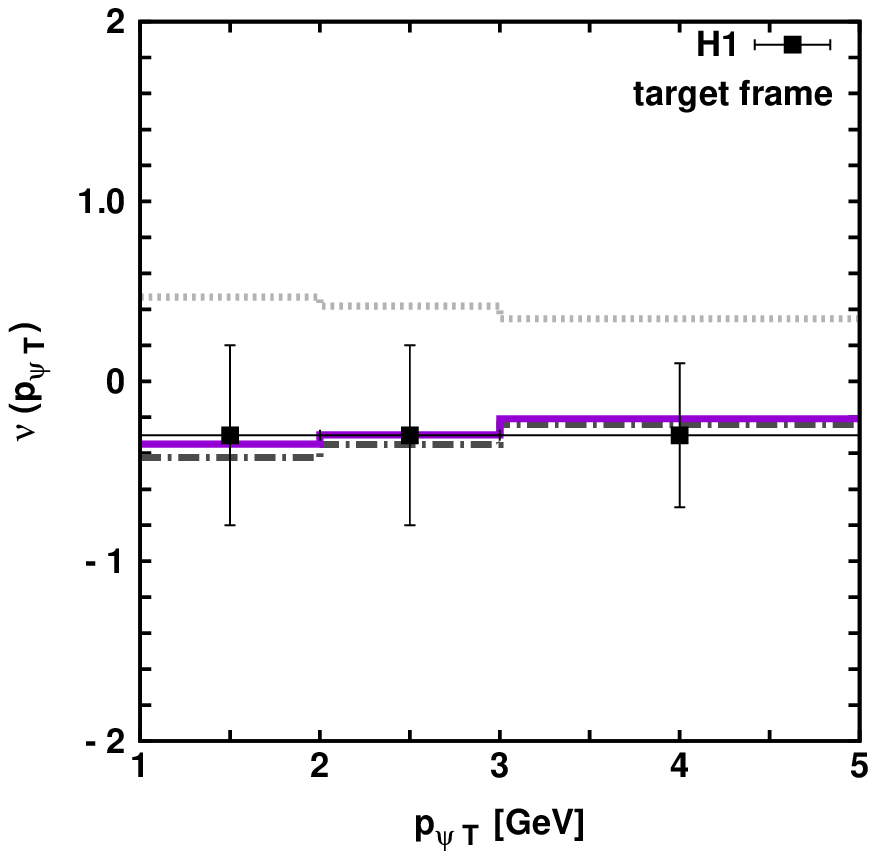, width = 8.1cm}
\epsfig{figure=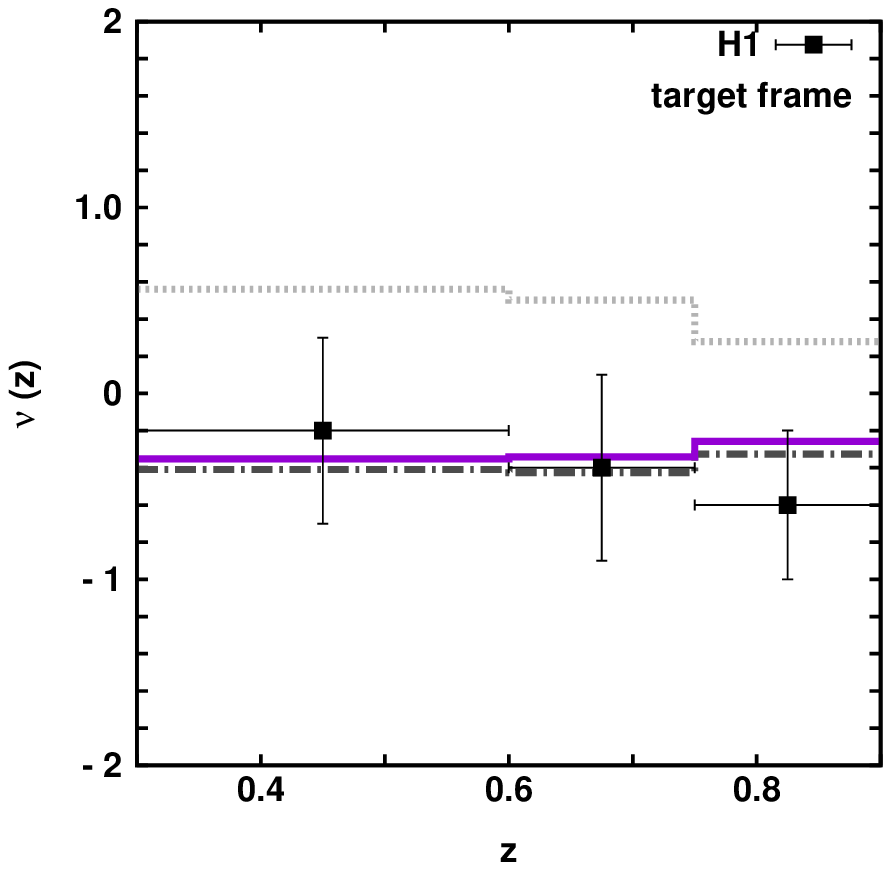, width = 8.1cm}
\caption{Polarization parameters $\lambda$ and $\nu$ as a 
functions of the $J/\psi$ transverse momentum and 
elasticity $z$ calculated in the target frame at
$0.3 < z < 0.9$, ${\mathbf p}_{\psi\,T}^2 > 1$~GeV$^2$ and $60 < W < 240$~GeV.
Notation of all histograms is the same as in Fig.~8.
The experimental data are from H1~\cite{28}.}
\end{center}
\label{fig15}
\end{figure}

\end{document}